\documentclass[10pt]{article}
\usepackage{amssymb,theorem,latexsym,authblk,amsmath,pstricks}
\usepackage{graphicx}
\usepackage{psfrag}
\usepackage{subfig}
\usepackage{epsfig}
\usepackage{epstopdf} 
\newcommand{\be}{ \begin{equation} }
\newcommand{\ee}{ \end{equation} }
\newcommand{\bea}{ \begin{eqnarray} }
\newcommand{\eea}{ \end{eqnarray} }
\newcommand{\ba}{ \begin{array} }
\newcommand{\ea}{ \end{array} }

\newcommand{\proof}{\noindent {\bf Proof. }}

\newcommand{\finpro}{\hfill $\Box$}
\newcommand{\ket}[1]{| #1 \rangle}
\newcommand{\bra}[1]{\langle #1 |}

\newcommand{\ketbra}[2]{| #1 \rangle \langle #2 |}

\newcommand{\e}{\mathrm{e}}

\newcommand{\I}{{\rm{i}}}
\newcommand{\D}{{\rm{d}}}
\newcommand{\E}{e}
\newcommand{\tr}{{\rm{tr}}}
\newcommand{\spec}{{\sigma}}

\def\integer{{\mathbb{Z}}}
\def\spinteger{{\mathbb{N}}^\star}
\def\real{{\mathbb{R}}}
\def\complex{{\mathbb{C}}}
\def\torus{{\mathbb{T}}}

\newtheorem{theo}{Theorem}

\newtheorem{prop}{Proposition}
\newtheorem{lemma}{Lemma}

\newtheorem{rem}{Remark}

\newcommand{\Span}{\operatorname{span}}

\newcommand{\ie}{i.e.}
\newcommand{\Zd}{\mathbb{Z}^d}

\newcommand{\CN}{{\mathbb{C}^N}}

\newcommand{\la}{\lambda}

\newcommand{\HP}{{H}_{P}}
\newcommand{\HPLa}{{H}_{P}^{\Lambda}}

\newcommand{\Hint}{{H}_{\rm int} }
\newcommand{\Hkin}{{H}_{\rm{hop}} }
\newcommand{\Htot}{{H}_{\rm{tot}}^{\Lambda} }
\newcommand{\HcP}{{\mathcal{H}}_P}
\newcommand{\HcB}{{\mathcal{H}}_B}

\newcommand{\observable}{{\cal B} ({\cal{H}}_P)}
\newcommand{\spinobs}{{\cal B}(\CN)}
\newcommand{\traceclass}{{\cal B}_1 ({\cal{H}}_P)}

\newcommand{\state}{{\cal S}_P}
\newcommand{\rhoP}{\rho_P}
\newcommand{\rhosl}{\rho_{\rm sl}}
\newcommand{\propa}[1]{{\cal Z}_{#1,\lambda}}

\newcommand{\propaLa}[1]{{\cal Z}_{#1,\lambda}^{\Lambda}}

\newcommand{\propaDyson}[1]{{\cal D}_{#1,\lambda}}
\newcommand{\propaDysonLa}[1]{{\cal D}_{#1,\lambda}^{\La}}
\newcommand{\ptlim}{\operatornamewithlimits{\rm{pt-lim}}}

\newcommand{\xv}{\underline{x}}

\newcommand{\lv}{\underline{l}}
\newcommand{\tv}{\underline{t}}

\newcommand{\tauv}{\underline{\tau}}

\newcommand{\Aa}{{\cal A}}
\newcommand{\Bb}{{\cal B}}

\newcommand{\Dd}{{\cal D}}

\newcommand{\Ii}{{\cal I}}

\newcommand{\Kk}{{\cal K}}
\newcommand{\Ll}{{\cal L}}

\newcommand{\Pp}{{\cal P}}

\newcommand{\Ss}{{\cal S}}
\newcommand{\Ttt}{{\cal T}}

\newcommand{\Vv}{{\cal V}}
\newcommand{\Ww}{{\cal W}}

\newcommand{\caA}{{\mathcal A}}
\newcommand{\caB}{{\mathcal B}}
\newcommand{\caC}{{\mathcal C}}

\newcommand{\caH}{{\mathcal H}}

\newcommand{\caK}{{\mathcal K}}
\newcommand{\caL}{{\mathcal L}}

\newcommand{\caU}{{\mathcal U}}

\newcommand{\str}{|}

\newcommand{\norm}{\|}
\newcommand{\bbZ}{\mathbb{Z}}

\newcommand{\links}{L}
\newcommand{\rechts}{R}
\newcommand{\La}{\Lambda}

\newcommand{\firstpairing}{\iota}

\title{Derivation of some translation-invariant Lindblad equations for a quantum 
Brownian particle}

\author[1]{Wojciech  De Roeck \footnote{email: wderoeck@thphys.uni-heidelberg.de}}
\affil{ \it Institut f\"ur Theoretische Physik, Universit\"at Heidelberg, 
Philosophenweg 16, 
D-69120 Heidelberg,  Germany}

\author[2,3]{Dominique Spehner \footnote{email: Dominique.Spehner@ujf-grenoble.fr}}

\affil[2]{\it Universit\'e Grenoble 1 and CNRS, Institut Fourier UMR5582, B.P. 74, 38402
Saint-Martin d'H\`eres, France}

\affil[3]{\it  Universit\'e Grenoble 1 and CNRS, Laboratoire de Physique et Mod\'elisation des
Milieux Condens\'es UMR5493,  B.P. 166, 38042 Grenoble, France}

\begin{document}

\maketitle

%\date{\today}

\begin{abstract}
We study the dynamics of a Brownian
quantum particle hopping  on an infinite lattice with a spin degree of freedom. 
This particle is coupled to free boson gases via a translation-invariant Hamiltonian 
which is linear in the creation and annihilation operators of the bosons.
We derive the time evolution of the reduced density matrix of the particle in the van Hove 
limit in which we also rescale the hopping rate. This corresponds to a situation in which  
both the system-bath interactions 
and the hopping between neighboring sites are small and they are effective on the same time scale. 
The reduced evolution is given by a translation-invariant 
Lindblad master equation which is derived explicitly. 
\end{abstract}

    \vspace{20pt}

%%%%%%%%%%%%%%%%%%%%%%%%%%%%%%%%%%%%%%%%%%%%%%%%%%%%%%%%%%%%%%
\section{Introduction}\label{sec: intro}
%%%%%%%%%%%%%%%%%%%%%%%%%%%%%%%%%%%%%%%%%%%%%%%%%%%%%%%%%%%%%%%

The irreversible dynamics of a quantum system coupled to infinite baths is often 
described by determining the time evolution of the reduced density matrix of the system, 
the latter being obtained by tracing out the bath degrees of freedom in the system\,+\,bath state.
Under certain approximations (including a  Born-Markov approximation), this density matrix
is the solution of a Lindblad master equation~\cite{Lindblad76,Kossakowski76}.
We are aware of three mathematically well-defined ways to derive such a Lindblad equation 
starting from the Hamiltonian dynamics of the system and 
baths~\cite{Spohn80,lebowitzspohn,alickiinvitation}: 
the weak coupling limit, the singular coupling limit, and the low density limit.
The weak coupling limit goes back to \cite{vanhove} and it was put on a rigorous footing in a 
series of papers by E.B.~Davies~\cite{Davies74,Davies76}. 
It consists in letting the system-bath coupling constant $\lambda$ going to zero and rescaling
time like $t = \lambda^{-2} \tau$, with $\tau>0$ fixed. 
This limit enforces the separation of time scales
\be \label{eq: timescales wc}
t_{S} \ll  t_R \;,\qquad t_B \ll t_R  \;,
\ee
where $t_S$ (sometimes called the ``Heisenberg time'') is
the time scale on which the system evolves in the absence of coupling with the baths,
$t_B$ is the  correlation time of the baths, and $t_R \approx t$ is  the time scale at which we 
describe the dynamics, that is, the time scale on which the system evolves under the coupling with the baths
(the ``relaxation time'' for systems
converging to stationary states).
The first time scale separation in (\ref{eq: timescales wc}) 
allows  to perform the rotating wave approximation. The second time scale separation allows for the Born-Markov
 approximation. The singular coupling limit
is the limit of delta-correlated baths, corresponding to
\be \label{eq: time scales singular}
t_{B} \ll t_{S} \; , \qquad t_B \ll t_{R}   \;.
\ee
Such a limit, which  is a quantum 
analog of the white noise limit for classical stochastic processes, 
has been  analyzed rigorously in~\cite{Lieb73,Gorini76,Palmer77}. 
It is physically meaningful in the limit of large bath temperature. 
Finally, we refer the reader to~\cite{Duemcke85} for a description of the low-density limit.  

The weak coupling, singular coupling, and low-density limits  have been applied and defined primarily 
for confined systems (typically atoms) 
coupled to free fermion or free boson baths.
There is a compelling reason for this in the case of the weak coupling limit: 
the Hamiltonian of a confined system has discrete spectrum, and therefore a well-defined time scale 
$t_S$ (given by the maximum of the inverse level spacings);
in contrast, extended systems may have continuous spectra, corresponding to arbitrarily slow 
processes in the uncoupled system dynamics, thus invalidating \eqref{eq: timescales wc} 
(see, however, \cite{taj} for a different approach).  A physical example of this is diffusion, 
where the relevant time scale is set by a spatial scale. In contrast, the singular coupling limit 
remains well-defined for extended systems, as one can guess from inspection 
of \eqref{eq: time scales singular} and as we will illustrate  in this article.
Note that  in the physics literature the dynamics of systems with arbitrarily large $t_S$ are often described 
by a Bloch-Redfield master equation with a time-dependent generator which is not of the Lindblad form
(see~\cite{Breuer} and references therein).
This equation is perturbative in the system-bath coupling but does not include a rotating wave approximation.

The derivation of the reduced dynamics of extended quantum systems is considerably more involved.
In \cite{erdos}, an
extended system is studied in the scaling limit
$t = \lambda^{-2} \tau$, $x = \lambda^{-2} \xi$, 
$\lambda \rightarrow 0$ (with $\tau>0$ and $\xi\in \real^d$ fixed)
 in which both the time $t$ and the position $x$ are rescaled.
The scaling of space is
 dictated by the scaling of time, since  on the microscopic scale the particle moves a distance of order
  $\lambda^{-2}$ in a time of order $\lambda^{-2}$.
In this limit, the resulting equation is a linear Boltzmann equation for the 
Wigner distribution of the particle.  
 This framework has been extended to describe decoherence in position space in \cite{Adamierdos} and  an essentially analogous result, with the weak coupling limit 
replaced by the low-density limit, was obtained in \cite{erdoseng}.

Let us also note that quantum systems coupled to infinite baths have been studied 
in the past fifteen year from another perspective. This approach, due to  
Jak${\check {\mathrm s}}$i\'{c} and Pillet~\cite{jaksicpillet1}, uses operator algebras and 
spectral analysis to describe the   
 dynamics of the system {\it and} bath  at large time and small but finite coupling constant
$\lambda$. 
For confined systems, we refer the reader to the lecture notes collected in 
\cite{attalreview} and the references therein. Extended systems have been analyzed
recently from a similar perspective in~\cite{deroeckdiffusionhighdimension, deroeckkupiainendiffusion}.
Another branch of activity on open quantum systems is the derivation of quantum stochastic equations, see for 
example~\cite{Gough05,Wojciech08}.

In this work, we consider an
extended system coupled to bosonic baths. We are interested in the dynamics of the 
reduced density matrix of this system
 at long times and for  weak  couplings. 
The system is a quantum particle moving on an infinite lattice $\Zd$, which 
has some internal degrees of freedom acting on a finite-dimensional Hilbert space.
In the simplest case, the Hamiltonian of the particle is 
the sum of the discrete Laplacian $-\Delta$ on 
$\ell^2 (\Zd)$ and of a self-adjoint Hamiltonian $S$ describing  the internal degrees of freedom.
One may also think of more general Hamiltonians coupling the position and 
internal degrees of freedom.
The particle interacts with free boson gases via a translation-invariant Hamiltonian,
assumed to be linear in the creation and annihilation operators of the bosons.
We consider the following scaling limit: 
(i) the time is rescaled as $t = \lambda^{-2} \tau$, where  $\lambda$
is the particle-boson coupling constant 
and   $\tau >0$ is fixed (ii)  the particle Hamiltonian is  rescaled as 
$H_P = - \lambda^2 \Delta + S$ (iii) one takes the limit
 $\lambda\rightarrow 0$.
This scaling combines the weak coupling and the singular coupling limits:  if
the translational degrees of freedom are frozen, it reduces to the 
weak coupling limit for the internal state,
whereas  if one ignores the internal degrees of freedom it amounts to a singular coupling limit for the
 motional state, as will be explained below 
(see~\cite{Spohn80,Palmer77}). In the latter case, however,  
the master equation is trivial. Indeed, 
by energy conservation the particle can only absorb or emit bosons with a vanishing frequency in the limit 
$\lambda \rightarrow 0$; since such bosons have also a vanishing momentum and the total momentum
is conserved, one has no momentum transfers between the particle
and the baths; thus the coupling to the baths has no effect on the particle, except possibly for
 decoherence in the momentum basis.
This is in fact the main reason why we consider a particle with 
internal degrees of freedom even though we are primarily concerned with its motion on the lattice. 
Note that since the hopping strength is of order $\la^2$, we do not need a space rescaling 
as in \cite{erdos, Adamierdos, erdoseng}. 

The above model allows for a tractable  rigorous analysis
in spite of the fact that we deal with a spatially extended system.
Our main result states that, in dimension $d\geq 2$,  the reduced density matrix of the 
particle converges in the aforementioned scaling limit to    
the solution of a Lindblad master equation which is determined explicitly.
This equation contains the physics of dissipative extended systems, in particular diffusion 
(whose analysis is, however, not treated in this work, 
we refer the reader to~\cite{deroeckdiffusionhighdimension,deroeckkupiainendiffusion} for results in this direction). Its derivation requires much less mathematical complications than in the works 
\cite{erdos, Adamierdos, erdoseng}.

Some related models have been studied 
in~\cite{Hellmich,hornbergervacchinireview,smirnevacchini,Bruneau2012}. In particular,
 Vacchini and coworkers 
considered in a series of non rigorous works~\cite{hornbergervacchinireview,smirnevacchini}  
similar models but in a low density limit; 
they argue that 
the evolution of the particle density matrix  (that is, not only 
of the associated Wigner transform) is governed in this limit by a Lindblad master equation. 
In~\cite{Bruneau2012}, drift and diffusion of  an electron moving on a one-dimensional 
lattice and submitted to a static electric field have been studied in a model in which the coupling to
the bath is simulated by repeated interactions with two level systems.
Finally, we point out that our model can be viewed as a continuous version of a 
dissipative quantum walk~\cite{Attal2012}.

The paper is organized as follows.   
We introduce the model  in Section~\ref{sec-model}, first at finite volume for the particle and  baths
and then by considering the infinite volume limit.   
Our results are presented and discussed in Section \ref{sec: results}, together with two important examples.
The last section  \ref{sec: proof} contains the proofs and some technical results are proven in the 
%two appendices. 
appendix.

%%%%%%%%%%%%%%%%%%%%%%%%%%%%%%%%%%%%%%%%%%%%%%%%%%%%%%%%%%%%%
\section{The model} \label{sec-model}
\subsection{The quantum particle} \label{sec-model_particle}
%%%%%%%%%%%%%%%%%%%%%%%%%%%%%%%%%%%%%%%%%%%%%%%%%%%%%%%%%%%%

Our model consists of a quantum particle on the lattice  $\Zd$ coupled to 
free boson fields. In this subsection and the three following ones, we
describe the model at finite volume\,\footnote{
In the mathematical literature on open quantum systems, it is common
to work  from the beginning with baths in the thermodynamical limit; the bath state and dynamics
are defined 
 with the help of (a representation of) an abstract CCR algebra
(see e.g.~\cite{attalreview}). We shall not use  
such an approach here and treat together the infinite lattice limit
for the particle and the thermodynamical limit for the bath.}. 
 We thus restrict the lattice to a finite hypercube with periodic boundary conditions
and consider  $\La = \mathbb{Z}^d/(2L \mathbb{Z})^d$  with $1 \leq L < \infty$, 
$d $ being the space dimension.  We will often  identify $\La$ with 
 $]-L,L]^d \cap  \Zd \subset \Zd$. 
The infinite volume limit $L\rightarrow \infty$ 
will be taken  in Subsection~\ref{sec: thermo limit}. 
The particle has translational degrees of freedom 
$x \in  \La $ and an internal degree of freedom 
$s =1,\ldots, N <\infty$, which may 
correspond to a 
spin or to an internal state of an atom or a molecule.
The Hilbert space of the particle, $\HcP^\La =\ell^2 (\La) \otimes \complex^N$, has finite
dimension $| \La | \times N$, where $| \La |= (2 L)^d$ is the cardinality of $\La$.
The particle Hamiltonian $\HPLa$ consists of a hopping term acting on 
$\ell^2 (\La)$ plus a term  governing the internal dynamics
 given by a self-adjoint operator $S$  acting on 
$\CN$,
\be \label{eq-HP}
\HPLa =  \lambda^{\alpha}  H_{\rm{hop}}^\La +  S
\ee
where we identified $S$ with $1_{\ell^2(\La)} \otimes S$.  
We have introduced in front of the hopping term
a small parameter $\lambda^{\alpha}$  playing the role of a hopping strength or of an inverse mass;
$\lambda >0$
will be chosen below to be the particle-boson coupling constant and 
$\alpha$ is a positive scaling exponent. Our most interesting result will correspond to $\alpha=2$
and $d \geq 2$.  Note that for $\alpha=\infty$ and $\la  <1$, \ie, $\lambda^\alpha=0$,
the translation 
degrees of freedom can be dropped altogether and the dynamics takes place on $\mathbb{C}^N$.
In the simplest setup, $\Hkin^\La$ is (up to a minus sign) the discrete Laplacian on 
$\ell^2 (\La)$,
\be \label{eq-discrete_Laplacian}
\Hkin^\La= - \Delta=  -  \sum_{x,y\in \La, |x-y|_\La =1}
\left( \ketbra{x}{y}  -  \ketbra{x}{x} \right) 
\otimes 1_{\complex^N}
\ee 
where $\{ \ket{x}; x \in \La\}$ is the canonical basis of $\ell^2(\La)$, 
 the Dirac notation $\ketbra{x}{y}$ refers to   
the operator $\psi \mapsto \psi(y) \ket{x}$ from $\ell^2 (\La)$ to $\ell^2 (\La)$, and 
  $
| x - y |_\La = \sum_{i=1}^d  \min_{k \in \integer}  | x_i - y_i + 2kL | 
$, i.e., we use periodic boundary conditions. 
In a more elaborate setup, $\Hkin^\La$ will be modified such that the propagation of the
 particle may couple to the internal state. 
An Hamiltonian that accommodates this idea 
is presented in Section~\ref{sec: setup nonequilibrium}.

An important property of our model is invariance under space
translations. These translations are represented on $\HcP^\La $ by
unitary operators $U^{\La}_{P}(x)$ 
defined by 
$U_{P}^\La (x) \ket{y} \otimes \ket{\phi} = \ket{x + y} \otimes \ket{\phi}$ for any 
$x,y \in \La$ and $\ket{\phi} \in \complex^N$. 
We state some conditions on $\Hkin^\La$, which are in particular satisfied by 
 the Hamiltonian \eqref{eq-discrete_Laplacian}.
{\it \begin{itemize}
\item[(A1)] 
 The hopping Hamiltonian $\Hkin^\La$ has  the form
 \be
 \Hkin^\La = \sum_{x \in \La}  U_{P}^\La (-x)  h_{\mathrm{hop}}  U_{P}^\La (x)  
= \sum_{x,y,z \in \La} \ketbra{y-x}{z-x} \bra{y} h_{\mathrm{hop}} \ket{z}
\label{eq: translation invariance}
\ee
where $h_{\mathrm{hop}}$ is 
a  $\Lambda$-independent self-adjoint operator on $\ell^2(\bbZ^d)\otimes \mathbb{C}^N$ satisfying 
\be \label{eq: finite range}
 \bra{y}  h_{\mathrm{hop}} \ket{z}   = 0 \qquad  \text{whenever} \, \,    |y|  >R \quad \textrm{or} \quad  |z|  >R 
\ee
for some $R<\infty$.   
In particular, this implies that $\Hkin^\La$ is translation-invariant, 
$U_{P}^\La (-x)   \Hkin^\La  U_{P}^\La (x) =  \Hkin^\La$ for any $x \in \La$, and has a finite 
 range independent of $\Lambda$. 
\end{itemize}
}
\noindent In the case of the discrete Laplacian \eqref{eq-discrete_Laplacian}, we have  
$h_{\mathrm{hop}}=(1/2) \sum_{\str z \str =1} 
( 2\str 0 \rangle \langle  0 \str -\str 0 \rangle \langle z \str-\str z \rangle \langle 0 \str) 
\otimes 1_{\complex^N}$.

By using  the Combes-Thomas estimate, 
which can be applied independently of $S$ since the latter operator does not act on the translation degrees 
of freedom, 
one can show that the finite range condition (\ref{eq: finite range}) implies the 
propagation bound  
\be \label{eq: propagation estimate}
\bigl\| \bra{x}  \e^{-\I  t \HPLa} \ket{y} \bigr\|
\leq \e^{\kappa \lambda^\alpha | t|} \e^{-|x-y|_\La}
\ee
 for some positive and $\La$-independent constant $\kappa$. Here and in what follows, $\| A \|$ denotes the 
operator norm of the operator $A$ (acting either on $\complex^N$, $\HcP^\La$, or another space);
  in (\ref{eq: propagation estimate}), the quantity
inside the  norm is a $N\times N$ matrix
 acting on the internal degrees of freedom of the particle.

%If $\La=\Zd$, then (\ref{eq: propagation estimate}) follows by the
%Combes-Thomas argument.  For finite $\La$, one can tile $\Zd$ with translates of $\La$ and then  \eqref{eq: propagation estimate} follows, too. 
%By translation invariance, it is enough to treat the case 
%$y=0$. This can be done by 
%introducing the operator
%$\Hkin^\La  = \e^{| X|} \Hkin^\La \e^{- | X|}$, where $|X|$ is the self-adjoint
%operator $|X| = \sum_{x \in \La} |x| \,\ketbra{x}{x}$. One then proves that  
%$\Hkin^\La $ is bounded for the operator norm by a $\Lambda$-independent 
%$c$ and this yields the result.   

%%%%%%%%%%%%%%%%%%%%%%%%%%%%%%%%%%%%%%%%%%%%%%%%%%%%%%%%%%%%%%%%
\subsection{The bosonic baths}
%%%%%%%%%%%%%%%%%%%%%%%%%%%%%%%%%%%%%%%%%%%%%%%%%%%%%%%%%%%%%

The particle is coupled to one or several bosonic baths labelled by $i \in I$
($I$ is a finite set). Let  $\torus^d$ be the $d$-dimensional torus, 
identified with the hypercube   $ ]-\pi, \pi]^d$.
Let $\La^*= (\pi/L) \Zd \cap \torus^d \setminus \{0\}$ be the dual of the 
lattice $\La$ after having removed the origin\,\footnote{
  In our finite-volume setup,
bosons with zero quasi-momentum $q=0$ do not play any role and should be
removed in order that all expressions in this section be well-defined.}.
The frequency 
$\nu_i(q) $ of a boson with a (nonzero) quantized momentum  $q \in \La^*$ is 
the value at $q$
of a function  
$\nu_i: q \in \torus^d \mapsto \nu_i(q) \in \real_{+}$ (dispersion relation of the  bath $i$). 
We assume that $\nu_i$ is continuous on $\torus^d$, 
$C^\infty$ on $\torus^d \setminus \{ 0\}$, and it satisfies
\be 
  \nu_i(q)> 0 \;\text{ for \;$q \not=0$.}
\ee
The Hilbert space of bath $i$ is the symmetric bosonic Fock space built 
on $\mathfrak{h}_i = \ell^2 (\La^* )$,
\be
\mathcal{H}_{\mathrm{B},i}^\La = \Gamma_s ( \mathfrak{h}_i ) =  
\mathbb{C}\,  \oplus \,   \mathfrak{h}_i \,  \oplus \,  ( \mathfrak{h}_i  \otimes_{\mathrm{s}} \mathfrak{h}_i ) 
 \,  \oplus \,  \cdots 
\ee
where $\otimes_{\mathrm{s}}$ stands for the symmetrized tensor product (see~\cite{Bratteli}  
for details and background).
The full bath space is then given by 
$\mathcal{H}_{B}^\La =\bigotimes_{i \in I}  \mathcal{H}_{B,i}^\La$.
The boson Hamiltonian $H_B^\La$, acting on $\mathcal{H}_B^\La$, is 
\be \label{eq-free_boson_Hamiltonian}
H_B^\La = \mathop{\sum}\limits_{i\in I}\sum_{q \in \La^* } 
\,  \nu_i(q)\,  a^\ast_{i,q} a_{i,q}  
\ee
where $a^\ast_{i,q}$ and $a_{i,q} $ are the  creation and 
annihilation operators for bosons with momentum $q$ in the bath $i$. We recall
that
$a^\ast_{i,q}$ and $a_{i,q} $ are unbounded operators 
on $\mathcal{H}_{B}^\La$,  acting trivially on $\mathcal{H}_{B,j}^\La$ with $j \not=i$, which  satisfy 
the canonical commutation relations $[a_{i,q}, a^\ast_{j,q'}] = \delta_{i,j} \delta_{q,q'}$ and
$[a_{i,q}, a_{j,q'}]=0$.  

%%%%%%%%%%%%%%%%%%%%%%%%%%%%%%%%%%%%%%%%%%%%%%%%%%%%%%%%%%%%%%%%
\subsection{Coupling between the particle and baths} \label{sec-model_coupling}
%%%%%%%%%%%%%%%%%%%%%%%%%%%%%%%%%%%%%%%%%%%%%%%%%%%%%%%%%%%%%

The particle and baths are coupled via a 
translation-invariant interaction Hamiltonian acting on 
$\mathcal{H}^\La = \HcP^\La \otimes \mathcal{H}_B^\La$,
\be  \label{eq-Hint_bis}
\Hint^\La 
=\frac{1}{\sqrt{| \La |}} \sum_{i\in I} \sum_{q \in \La^*}  
 \,g_{0,i}(q)  W_{i} \otimes \e^{\I q \cdot X }\otimes (   a^\ast_{i,q}   + a_{i,-q}  )
\ee
where $X$ is the position operator 
(acting on $\ell^2 (\La )$ as a multiplication by $x$   
 and acting trivially on $\complex^N$),
$W_{i}$ is an Hermitian $N\times N$-matrix that models the 
interaction with 
the internal degree of freedom, and   
$g_{0,i}(q)$ are momentum-dependent coupling constants. 
Having in mind the thermodynamical limit which will be considered below, 
we assume that $g_{0,i}(q)$ are the values at the quantized 
momenta $q \in \La^*$ of some continuous functions 
$g_{0,i}:\torus^d \mapsto \complex$.  These functions are 
called the form factors in the sequel. They must satisfy  
$g_{0,i}(q)=\overline{g_{0,i}(-q)}$,
$q \in \torus^d$, in order that $\Hint^\La$ be self-adjoint. 
Introducing the field operators
\be \label{eq-field_op}
\Phi^{\La}_i(\varphi) =  | \Lambda |^{-1/2}\sum_{q \in \La^* } \, \left(\varphi (q)\,a_{i,q}^\ast + \overline{\varphi (q)}\,a_{i,q}\right)
\ee
for 
$\varphi \in \ell^2 (\La^*)$,
one may rewrite $\Hint^\La$ as
\be \label{eq-Hint}
\Hint^\La 
= 
\sum_{i\in I} 
\sum_{x \in \La} W_{i} \otimes \ketbra{x}{x} \otimes \Phi^{\La}_i(g_{x,i})
\ee
where we have set
\be \label{eq-g_i,x}
g_{x,i}(q) = \e^{\I q \cdot x} g_{0,i}(q) \;.
\ee

Up to the freedom in the form factors $g_{0,i} (q)$, the choice of the Hamiltonian  
(\ref{eq-Hint_bis}) is dictated by 
the requirement that it must be invariant under 
space translations and linear in the creation and annihilation operators of the bosons. 
Space translations  are represented 
on the bosonic Fock space of the bath $i$ by unitary operators $U^{\La}_{i}(x)$ satisfying $U^{\La}_{i}(-x) a_{i,q} U^{\La}_{i}(x)
 = \e^{\I q \cdot x} a_{i,q}$. 
 One easily checks that for any $x \in \integer^d$,
\be
U^\La(- x) \Hint^\La U^\La (x) = \Hint^\La
\ee
with $U^\La (x) =U_P^\La (x) \otimes \bigotimes_{i \in I} U^{\La}_i (x)$.  
For instance, electrons in solids are coupled to low-energy acoustic 
phonons via
an Hamiltonian of the form (\ref{eq-Hint_bis}), see~\cite{Ziman,Ridley}.

The total Hamiltonian of the coupled system, acting on $\mathcal{H}^\La$, is
\be \label{eq-H_P_bis}
\Htot = ( \lambda^\alpha \Hkin^\La + S ) \otimes 1_{\HcB^\La}  
+  1_{\HcP^\La} \otimes H_B^\La +\lambda \Hint^\La   
\ee
where we have introduced the dimensionless coupling constant $\lambda$ in front of 
$\Hint^\La$.   Using $\nu_i(q)> 0$ and the finiteness of $\La$,  one can apply the Kato-Rellich theorem to conclude that 
$\Htot$ is self-adjoint on the domain of $H_B^\La$.
%
%The condition
%\be  
%  \sum_{q \in \La^*} \frac{| g_{0,i} (q)|^2 }{\nu_i (q)}   
% < \infty, \qquad  i \in I
%\ee
%% 
%ensures that 
%$\Hint^\La$ is relatively bounded with respect to $H_B^\La$ and 
%thus that 

%%%%%%%%%%%%%%%%%%%%%%%%%%%%%%%%%%%%%%%%%%%%%%%%%%%%%%%%%%%%%%%
\subsection{Initial state} \label{sec: initial state}
%%%%%%%%%%%%%%%%%%%%%%%%%%%%%%%%%%%%%%%%%%%%%%%%%%%%%%%%%%%%%%%%%

We assume that the particle and bosons are initially 
in a product state $\rhoP^\La  \otimes \rho_B^\La$, where 
$\rhoP^\La$ is the  initial density matrix of the particle and 
$\rho_B^\La$ the initial density matrix of the bosons (\ie,  $\rhoP^\La$ and 
$\rho_B^\La$ are positive operators on $\HcP^\La$ and $\HcB^\La$ with trace one).
We now specify our assumptions on the boson initial density matrix $\rho_B^\La$. 
To this end, we consider  the $n$-point correlation functions
\be
\tr \bigl(  \rho_B^\La \,
 a_{i_1,q_1}^{\#_1} a_{i_2,q_2}^{\#_2} \cdots a_{i_n,q_n}^{\#_n}  
\bigr)     \label{eq: first correlation function}
\ee
where $a_{i,q}^{\#}$ stands for $a_{i,q}$ or $a_{i,q}^{\ast}$. 
{\it \begin{itemize}
\item[(B1)] The bath density matrix $\rho_B^\La$ is translation-invariant and stationary with 
respect to the free 
dynamics generated by $H_B^\La$: 
\be
  U_i^\La (x)  \rho_B^\La U_i^\La (-x) = \e^{-\I t H_B^\La} \rho_B^\La  \e^{\I t H_B^\La} = \rho_B^\La 
\quad \text{ for any 
$x\in \La$, $i \in I$, and $t \in \real$.}
\ee
\end{itemize}
}
{\it \begin{itemize}

\item[(B2)]  $\rho_B^\La$ is quasi-free. That is, the correlation functions \eqref{eq: first correlation function} 
exist for any $q_1,\cdots, q_n \in \La^*$, they vanish if the number of creators is distinct from the number 
of 
annihilators\;\footnote{This property is sometimes called gauge invariance  since it follows from the invariance of the state under the gauge transformation $  a_{i,q} \to  \e^{\I \theta} a_{i,q}  $.}
 (in particular, if $n$ is odd) and they satisfy the following Gaussian property 
(also known as Wick's identity):  for $n$ even,  
\be \label{eq-Wick}
 \tr
\bigl(  \rho_B^\La \,
 a_{i_1,q_1}^{\#_1}  a_{i_2,q_2}^{\#_2} \cdots a_{i_n,q_n}^{\#_n} 
\bigr) 
= 
\sum_{{\rm{parings}}\;\underline{\pi}\;{\rm{of}}\;  (1,\cdots, n) } \;
\prod_{m=1}^{n/2}  
 \tr
\bigl(  \rho_B^\La \,
a_{i_{\firstpairing_m}, q_{\firstpairing_m}}^{\#_{\firstpairing_m}}
 a_{i_{\sigma_m},q_{\sigma_m}}^{\#_{\sigma_m}}  
\bigr) \;.
\ee
The sum in (\ref{eq-Wick}) runs over all pairing of $(1,\cdots,n)$, that is, over all sets 
$\underline{\pi}=\{ (\firstpairing_1,\sigma_1),\cdots, (\firstpairing_{n/2},\sigma_{n/2})\}$ of $n/2$ pairs of distinct indices such that
 $1=\firstpairing_1<\firstpairing_2<\cdots <\firstpairing_{n/2}$, $\firstpairing_m < \sigma_m$ for any $m=1,\cdots,n/2$, and
$\{ \firstpairing_1,\cdots,\firstpairing_{n/2}\} \cup \{ \sigma_1,\cdots ,\sigma_{n/2}\}=\{ 1, \cdots , n\}$. 

\item[(B3)] $\rho_B^\La= \otimes_{i \in I} \rho_{B,i}^\La$  is a product of 
  quasi-free density matrices $\rho_{B,i}^\La $ on $\mathcal{H}_{B,i}^\La$. In particular, this implies 
that  the two-point correlation function
$ \tr(  \rho_B^\La \, a_{i,q_1}^{\#_1}  a_{j,q_2}^{\#_2})$ vanishes if $i \neq j$. 
\end{itemize}
}

Note that, according to  assumptions {\it (B1)} and {\it (B2)}, 
$ \tr(  \rho_B^\La \, a_{i,q_1}^{\#_1} a_{j,q_2}^{\#_2})=0$ also vanishes if $q_1 \not= q_2$.
Assumption {\it (B3)} means that 
the baths are not correlated initially.  
Assumptions {\it (B1-B3)} imply that $\rho_B^\La$ is completely determined 
by the set $\{ \zeta_i^\La (q) ; q \in \Lambda^* \}$ of  
occupation numbers 
$\zeta_i^\La (q) = \tr ( \rho_B^\La \, a_{i,q}^\ast a_{i,q}) \in \real_{+}$ of bosons  with momentum 
$q$ in bath $i$. 
In particular, 
\be \label{eq: initial state}
 \tr
\bigl(  \rho_B^\La \,  \Phi^{\La}_i(\varphi_1)
\Phi^{\La}_i (\varphi_2) \bigr)  
 = \frac{1}{|\La |}
\sum_{q\in \La^*}  \, \left( \zeta_i^\La (q) \varphi_1 (q) \overline{\varphi_2 (q)} 
  +  (1+\zeta_i^\La(q)) \overline{\varphi_1 (q)} {\varphi_2 (q)}   \right)
\ee
for any $ \varphi_1,\varphi_2 \in \ell^2 (\La^*)$. 
When taking the thermodynamic limit we will need the additional hypothesis:
{\it \begin{itemize}
\item[(B4)] $\zeta_i^\La (q)$ are the values at the quantized 
momenta $q \in \La^*$ of some continuous function 
$\zeta_{i}:\torus^d \setminus \{0\} \to \real_{+}$ 
such that $ | g_{0,i} |^2 \zeta_i  \in L^1 (\torus^d )$. 
\end{itemize}
}

The prime example of an initial state satisfying  {\it (B1-B3)} is  
\be
\rho_B^\La  = \bigotimes_{i \in I} \rho_{\beta_i,i}^{\La} \quad \text{ with } \quad 
\rho_{\beta_i, i}^{\La}= \frac{\E^{-\beta_i H_{B,i}^\La}}{\tr ( \E^{-\beta_i H_{B,i}^\La})}
\ee
the Gibbs state at inverse
temperature $\beta_i >0$. This situation corresponds to 
a particle coupled to thermal baths (which may have different temperatures).
Then
$\zeta_i (q) = ( \e^{\beta_i \nu_i(q)} -1 )^{-1}$ is the Bose-Einstein distribution.  
If the form factors are such that $| g_{0,i}|^2 /\nu_i \in L^1 (\torus^d)$ then
the last assumption {\it (B4)} holds true. 

%%%%%%%%%%%%%%%%%%%%%%%%%%%%%%%%%%%%%%%%%%%%%%%%%%%%%%%%%%%%%%%%%%
\subsection{Thermodynamic limit} \label{sec: thermo limit}
%%%%%%%%%%%%%%%%%%%%%%%%%%%%%%%%%%%%%%%%%%%%%%%%%%%%%%%%%%%%%%%%%%%

To observe irreversible phenomena, we have to consider the baths at the thermodynamic limit, that is, 
send $\La \nearrow \Zd$ 
keeping the boson densities fixed. 
By $\Lambda \uparrow \mathbb{Z}^d$ we mean the limit $L \rightarrow \infty$, $L$ being the size of the
hypercube $\Lambda$; in this limit the motion of the particle takes place on the infinite lattice $\mathbb{Z}^d$.

Let $\Hkin$ be the bounded self-adjoint operator on $\HcP= \ell^2 (\Zd) \otimes \complex^N$
defined by $\bra{x} \Hkin^\La \ket{y} \rightarrow  \bra{x} \Hkin \ket{y}$
as $\La \nearrow \Zd$ for any $x,y \in \Zd$. This Hamiltonian is given formally by
$\Hkin = \sum_{x \in \Zd} U_P (-x) h_{\rm hop} U_P (x)$. 
It  describes the hopping of the particle between the lattice sites
in the infinite volume limit. 
We identify all operators on $\ell^2 (\La) \otimes \complex^N$ as finite-rank operators on 
$\ell^2 (\Zd ) \otimes \complex^N$.
By the finite range condition {\it (A1)}, 
$\bra{x} \Hkin \ket{y}=0 $ for  $|x-y| > 2 R$ and $\Hkin^\La \rightarrow \Hkin$ strongly as $\La \nearrow \Zd$. 
Since  $\Hkin^\La$ (and thus $\Hkin$) are bounded, it then follows, e.g.\ by the Duhamel formula, that
\be \label{strong_convergence_unitary_free_ev}
\e^{-i t \HP^\La} \rightarrow \e^{-i t \HP} 
\quad 
\text{ strongly as $\La \nearrow \Zd$} 
\ee
with $\HP = \lambda^\alpha \Hkin + S$.

In what follows, we will
denote by $\observable$ (respectively $\traceclass$) 
the Banach space  of  bounded (trace-class) operators 
on $\HcP$ 
(recall that the norm on  $\traceclass$  is the trace norm
$\| A  \|_1 = \tr ( | A |)$), and by $\state$ the 
convex cone of density matrices on $\observable$ (\ie,
positive operators in $\traceclass$ with  trace one).
We must clearly assume that the finite volume initial state of the particle converges 
as $L \rightarrow \infty$ in the trace-norm topology.
{\it
\begin{itemize} 
\item[(A2)] $\rho^\La_{P}   \mathop{\rightarrow}\limits_{\La \nearrow \Zd}  
\rho_P$ in $\traceclass$. 
\end{itemize}
}
 It is easy to show from the commutation relations of the 
$a_{i,q}^\#$'s that the field operator (\ref{eq-field_op})
evolves under the Hamiltonian (\ref{eq-free_boson_Hamiltonian}) as  
\be \label{eq-evolved_field_op}
\e^{i t H_B^\La} \Phi_i^\La (\varphi ) \e^{-i t H_B^\La} = \Phi_i^\La ( \e^{i t \nu_i} \varphi )\;.
\ee
The following space-and-time  bath correlation functions
\be \label{eq: correlation function real definition} 
f^{\La}_i(x,y;t,s)
 =  \tr
\bigl(  \rho^{\La}_B 
 \Phi^{\La}_i( \e^{i t \nu_i } g_{x,i} )
 \Phi^{\La}_i ( \e^{i s \nu_i } g_{y,i})  \bigr)
\ee
will play an important role in what follows. 
By translation invariance and stationarity of the bath initial state $\rho_B^\Lambda$
(assumption {\it (B1)}),
$f^{\La}_i(x,y;t,s)= f^{\La}_i(x-y,t-s)$
 only depends on the position difference $x-y$ and time
difference $t-s$, where, according to (\ref{eq: initial state}),
\be \label{eq-correl_function}
f^{\La}_i(x,t) = 
\frac{1}{|\La|} \sum_{q \in \Lambda^*} \str g_{0,i}(q)\str^2 
\left(  \zeta_i^\La (q)  \e^{i q \cdot x} \e^{i t \nu_i (q)}  + (1+ \zeta_i^\La (q))  \e^{-i q \cdot x} 
\e^{-i t \nu_i (q) }    \right) \;.
\ee 
 By assumption {\it (B4)}, $f^{\La}_i(x,t)$ 
converges  as  $\La \nearrow \Zd$ to 
\be \label{eq-assumption} 
f_i(x,t) = \int_{\torus^d} \frac{\D^d q}{(2\pi)^d} \, \str g_{0,i}(q)\str^2 
\left(  \zeta_i (q)  \e^{i q \cdot x} \e^{i t \nu_i (q)}  + (1+ \zeta_i (q))  \e^{-i q \cdot x} 
\e^{-i t \nu_i (q)  }    \right) 
\ee
 uniformly in $t$ (recall that $\nu_i$ is bounded).

%%%%%%%%%%%%%%%%%%%%%%%%%%%%%%%%%%%%%%%%%%%%%%%%%%%%%%%%%%%%%%%%%%
\subsection{Reduced density matrix of the particle} 
%%%%%%%%%%%%%%%%%%%%%%%%%%%%%%%%%%%%%%%%%%%%%%%%%%%%%%%%%%%%%%%%%%%

The reduced density matrix  $\rhoP^{\La} (t)$ of the particle 
at time $t\geq 0$  is
the partial trace over $\mathcal{H}_B^\La$ of the time-evolved  density matrix
of the total ``particle\,+\,bosons'' system,  
\be \label{eq-definition_reduced_evolution}
 \rho^{\Lambda}_{P} (t)
 = \tr_{B} \Bigl(   \e^{-\I t \Htot}   \rhoP^\La  \otimes \rho_B^\La \,
\e^{\I t \Htot} \Bigr)
\;.
\ee
The following proposition states that it  is well defined
in the thermodynamic limit 
under the  assumptions described in the preceding subsections.

%%%%%%%%%%%%%%%%%%%%%%%%%%%%%%%%%%%%%%%%%%%%%%%%%%%%%%%%%%%%%%%%%%%%%%%%%%%%%%5
\begin{prop}\label{lem: thermo limit}
Assume that (A1-A2) and (B1-B4)
are satisfied. Then for each $t \geq 0$ and $\lambda > 0$, the reduced density 
matrix \eqref{eq-definition_reduced_evolution} converges as $L \rightarrow \infty$,
\be \label{eq-thermodynamical_limit}
\rho^{\Lambda}_{P} (t) 
%= 
%\tr_B  \bigl( \e^{-\I t \Htot}   \rhoP^{\La}  \otimes \rho^{\La}_B \e^{\I t \Htot} 
% \bigr) 
\quad \mathop{\rightarrow}\limits_{\La \nearrow \Zd}  
\quad \propa{t} (\rhoP)   
\ee
in the trace-norm topology,
where $ \propa{t}$ is a completely and positive trace-preserving
map acting on $\traceclass$. 
\end{prop}
%%%%%%%%%%%%%%%%%%%%%%%%%%%%%%%%%%%%%%%%%%%%%%%%%%%%%%%%%%%%%%%%%%%%%%%%%%%%%%%

This proposition will be proven in section~\ref{sec_proof_thermodyn_limit}.

%%%%%%%%%%%%%%%%%%%%%%%%%%%%%%%%%%%%%%%%%%%%%%%%%%%%%%%%%%%%%%%%%%%%%
\section{Results and discussion} \label{sec: results}
 %%%%%%%%%%%%%%%%%%%%%%%%%%%%%%%%%%%%%%%%%%%%%%%%%%%%%%%%%%%%%%%%%%%%%

%%%%%%%%%%%%%%%%%%%%%%%%%%%%%%%%%%%%%%%%%%%%%%%%%%%%%%%%%%%%%%%%%%%%%%%%%%%%%%
\subsection{The scaling limit} \label{sec: scaling limit}
%%%%%%%%%%%%%%%%%%%%%%%%%%%%%%%%%%%%%%%%%%%%%%%%%%%%%%%%%%%%%%%%%%%%%%%%%%%%%%

 To obtain rigorously a kinetic equation for $ \rho_{P} (t)   =   \propa{t}(\rhoP) $, we
 perform a van Hove limit by setting
 $t =\lambda^{-2} \tau$
and letting $\lambda \rightarrow 0$
 while keeping the rescaled time $\tau>0$ fixed.
In the interaction picture with respect to the internal Hamiltonian $S$, 
the reduced density matrix of the particle is in the scaling limit 
\be \label{eq: limit of main result}
\rhosl(\tau) = 
\mathop{\lim}\limits_{\lambda \rightarrow 0,  t =\lambda^{-2} \tau \rightarrow \infty}     
 \e^{\I t S} \propa{t} (\rhoP)    \e^{-\I t S}
 \ee
(since we never consider objects on the bath space in this limit we write  
$\rhosl(\tau)$ instead of $\rho_{P, \mathrm{sl}} (\tau)$).
Note that the infinite volume limit (\ref{eq-thermodynamical_limit}) 
has been taken first, before letting $\lambda\rightarrow 0$.
Our main result is the existence and characterization of the limit 
(\ref{eq: limit of main result}).
Recall that $\lambda$ appears both in front of the interaction Hamiltonian $H_{\rm int}$ 
and of the hopping term $\Hkin$ in the total Hamiltonian (\ref{eq-H_P_bis}).
Hence  hopping between the lattice
sites goes to zero as  $\lambda \rightarrow 0$ and 
the motion induced by the Hamiltonian $\Hkin$ becomes effective only at large times 
$t \approx \lambda^{-\alpha} \tau$.
It is then intuitively clear that for $\alpha>2$ the hopping will be absent in our scaling limit, whereas
for $\alpha=2$ both hopping and dissipative effects due to boson absorptions and emissions
should be contained in the kinetic equation for $\rhosl(\tau)$.

Before stating the result, let us introduce some notation.
In the following, $\{ A , B\}=A B + B A$ denotes the anticommutator of two 
operators $A$ and $B$ on $\HcP$,
$\{ \ket{s}; s=1,\cdots, N\}$ is the 
orthonormal basis of $\complex^N$ diagonalizing the internal Hamiltonian $S$, and 
$E_s \in \spec (S)$ are the eigenvalues of $S$,  that is, $S \ket{s}= E_s \ket{s}$.
For any Bohr frequency $\omega \in \spec ([S,\cdot] )= \spec (S) - \spec(S)$,
we define the $N \times N$ matrix
\be
W_{i,\omega} 
 = 
  \sum_{s,s'=1,\cdots ,N }   \delta_{E_s-E_{s'}, \omega } \, 
\bra{s} W_{i}   \ket{s'} \ketbra{s}{s'}  
\ee
and the \emph{spectrally averaged} hopping Hamiltonian
\be \label{eq-spectrally_av_hopping_Hamil}
H^{\natural}_{\rm{hop}} =   \sum_{s,s'=1,\cdots ,N}       \delta_{E_s,E_{s'} }  \,     \bra{s} H_{\rm{hop}} \ket{s'}   \ketbra{s}{s'} 
\ee
where $\delta_{a,b}$ is the Kronecker delta symbol (equal to unity if $a=b$ and zero 
otherwise). Finally, let us recall that a quantum dynamical semigroup 
(QDS) on $\traceclass$
is a semigroup  $(\Ttt_\tau)_{\tau \geq 0}$ of completely positive maps  
$\Ttt_\tau: \traceclass \rightarrow \traceclass$ preserving the trace and such that
$\tau \in \real_{+}\mapsto \Ttt_\tau$ is $\ast$-weakly continuous. 
Lindblad~\cite{Lindblad76} has derived 
the general form of the generators of norm-continuous QDS (see also \cite{Kossakowski76} and  
an extension to unbounded generators in \cite{Davies79}).  

%%%%%%%%%%%%%%%%%%%%%%%%%%%%%%%%%%%%%%%%%%%%%%%%%%%%%%%%%%%%%%%%%%%%%%
\begin{theo} \label{thm: main}
Let assumptions (A1-A2) and (B1-B4) be satisfied and let $\alpha \geq 2$. 
Assume moreover that the infinite-volume correlation functions $f_i(x,t) $  satisfy
\be \label{eq-integrability_assumption}
\lim_{n \rightarrow \infty} \frac{1}{n} \int_0^\infty \D t \,
 \sup_{x \in \Zd, i\in I} \, | f_i(x,t)| \e^{-\frac{| x|}{n} }
  = 0 \;.
\ee
Then  for any $\tau>0$ and any $\rhoP \in \state$,
the limit \eqref{eq: limit of main result} exists in the trace-norm
topology
 and is equal to $\rhosl (\tau) = \e^{\tau \Ll^\natural} \rhoP$, where 
$(\e^{\tau \Ll^\natural})_{\tau\geq 0}$ 
is a norm-continuous quantum dynamical semigroup on $\traceclass$. If 
$\alpha=2$ the generator $\Ll^\natural$ of this semigroup is given by  
\begin{eqnarray} \label{eq-Lindblad_generator}
\nonumber
\Ll^\natural (\rho)
& = &  - \I \bigl[ H^{\natural}_{\rm{hop}} + \Upsilon, \rho \bigr]   + 
 \sum_{\omega \in \spec ([\cdot, S]) } \sum_{i\in I} 
   \Biggl(  \sum_{x,y\in \Zd}  c_i(y-x,\omega) 
    W_{i,\omega}\otimes \ketbra{x}{x} \, \rho\, 
W_{i,\omega}^\ast \otimes \ketbra{y}{y}
\\[3mm]
    &&
-\frac{c_i(0,\omega)}{2}   \Bigl\{  W_{i,\omega}^\ast  W_{i,\omega} \otimes 1_{\ell^2(\Zd)}
\,,\, \rho \Bigr\} \Biggr)
\end{eqnarray}
where
\be \label{eq-level_shift}
\Upsilon  =  \sum_{\omega \in \spec ([\cdot, S])}   \sum_{i\in I}
\Im \left\{ \int_{0}^{\infty} \D t \, f_i(0,t)    \e^{-i t \omega} \right\} \,
 W_{i,\omega}^\ast  W_{i,\omega} \otimes 1_{\ell^2(\Zd)} 
\ee
and $c_i(x,\omega)$ is the time Fourier transform of $f_i(x,t)$, 
\be \label{eq: expression c}
 c_i(x,\omega)  = 
 \int_{\mathbb{R}} dt  f_i(x,t)  \e^{-i t\omega} \;.
\ee
If $\alpha>2$, $\Ll^\natural$ is given by the same expression as in 
(\ref{eq-Lindblad_generator}) but without the term $-\I [ \Hkin^{\natural} ,\rho]$. 
\end{theo}

To see that $\Ll^\natural$ in (\ref{eq-Lindblad_generator})  
has the 
Lindblad form (and thus that $(\e^{\tau \Ll^\natural})_{\tau\geq 0}$ 
is a QDS),  we first rewrite
\be 
\Ll^\natural (\rho) = -\I [ H^\natural_{\rm{hop}} + \Upsilon, \rho] +\mathfrak{A}(\rho)- \frac{1}{2} \{ \mathfrak{A}^{\star}(1),\rho \}   \label{eq: manifest lindblad form}
\ee
where the map $\mathfrak{A}$ abbreviates the term involving a sum over $x$ and $y$ 
in the right-hand side of 
\eqref{eq-Lindblad_generator} and $\mathfrak{A}^{\star}$ is its adjoint with respect  to the trace, i.e., 
$\tr(\mathfrak{A}(\rho) A)=\tr(\rho\mathfrak{A}^*(A) ) $. 
We note  that $c_i(x,\omega) $ is of positive type in the $x$-variable  
(this follows from the fact that $f_i(x,t)$ is a correlation function, 
see \eqref{eq: correlation function real definition})
and therefore it is the Fourier transform of a positive measure $\widehat{c}_{i} (\D q, \omega)$ on $\torus^d$:
\be \label{eq-FT_c}
c_i ( x, \omega) 
 =  \frac{1 } { (2\pi)^d} 
  \int_{\torus^d}   \widehat{c}_i (\D q, \omega ) \,\e^{-\I q \cdot x} \;.
\ee  
This shows that $\mathfrak{A}$ has the Kraus form
\be  \label{eq-LL_Lindblad_form}
\mathfrak{A}(\rho) = \sum_{\omega \in \sigma( [\cdot, S])} \sum_{i\in I} 
 \int \widehat{c}_i ( \D q, \omega) V_{i,\omega} (q) \,\rho \, V_{i,\omega} (q)^\ast
\ee
with
$V_{i,\omega} = (2\pi)^{-d/2}  W_{i,\omega} \otimes \e^{i q \cdot X}$. Thus $\mathfrak{A}$
is a completely positive map~\cite{Kraus70}. 
Moreover, $\mathfrak{A}$ is bounded  on $\traceclass$ because for any $\rho \in \Bb (\HcP)$, $\rho \geq 0$,
\be
\norm\mathfrak{A}(\rho) \norm_1 = \tr \bigl( \mathfrak{A}(\rho)\bigr)   \leq 
\sum_{\omega \in \spec ([\cdot, S]) } \sum_{i\in I} 
 c_i(0,\omega)  \norm W_{i,\omega} \norm^2 \, \norm \rho \norm_1 
\ee
and $c_i(0,\omega) $ is finite by the integrability of the correlation function $f_i (0,t)$
(assumption~(\ref{eq-integrability_assumption})). 
Since  also $\mathfrak{A}^\ast(1)$ and $H_{\rm{hop}} + \Upsilon$ are bounded operators on $\mathcal{H}_P$, 
it follows that $\Ll^\natural $ is bounded on $\traceclass$.   
This boundedness and the complete positivity of $\mathfrak{A}$ imply that 
the operator $\Ll^\natural$ in (\ref{eq: manifest lindblad form}) generates a norm-continuous QDS~\cite{Lindblad76}.

Another way of phrasing Theorem~\ref{thm: main} is to say that the rescaled density matrix $ \rhosl(\tau)$ satisfies the 
\emph{Bloch-Boltzmann} master equation
$$
\frac{\D }{\D \tau} \rhosl (\tau)  = \Ll^\natural \bigl( \rhosl (\tau) \bigr)\;.
$$
Note that $\Ll^\natural$ commutes with $\I [ S,\cdot]$,
a generic fact for generators obtained via a weak coupling limit~\cite{Davies74,Spohn80}.
The self-adjoint operator 
$\Upsilon$ in (\ref{eq-level_shift}) acts trivially on $\ell^2 (\Zd)$ and 
commutes with $S$; it represents the energy shifts of the particle
due to its coupling 
with the bosons (Lamb shifts). 
In the following sections~\ref{sec: setup equilibrium} and \ref{sec: setup nonequilibrium}, 
we unwrap the form of the generator $\Ll^\natural$ 
in two different situations and discuss the physical 
phenomena described by the corresponding master equation.

The major technical assumption of Theorem~\ref{thm: main} is the 
integrability condition (\ref{eq-integrability_assumption}) on the boson correlation functions. 
This assumption should be compared to the analogous condition for confined systems~\cite{Spohn80,Davies74,Davies76}, 
\ie,  the integrability of  the correlation function
(\ref{eq-assumption}) for $x=0$.
An explicit computation and the use of a stationary phase argument
performed in Appendix~\ref{app-C} yields:

%%%%%%%%%%%%%%%%%%%%%%%%%%%%%%%%%%%%%%%%%%%%%%%%%%%%%%%%%%%%%%%%%%%%%%5
\begin{prop} \label{prop-decay_correl_funct}
Let us assume that the form factor $g_{0,i}$ has a support contained in the open ball 
$\{ q \in \torus^d ; |q|<\pi \}$,  that $|g_{0,i}|(q)$ and $\zeta_i(q)$ depend only on the modulus $|q|$
of $q$, and that the bosons of bath $i$  have a linear dispersion relation
$\nu_i (q)= |q|$ on the support of $g_{0,i}$. 
Furthermore, let the functions $\psi_{i,+} (|q|) = |g_{0,i} (q) |^2 \zeta_i (q)$
and $\psi_{i,-} (|q|) = |g_{0,i} (q) |^2 (1+ \zeta_i (q))$ belong to $C^2(]0,\pi])$ and 
\be
|q|^{\min\{d-3,\frac{d-1)}{2} \} } \psi_{i,\pm} (|q|)
\quad , \quad
 |q|^{d-2} \psi_{i,\pm} ' (|q|) 
\quad , \quad 
|q|^{d-1} \psi_{i,\pm} '' (|q|)
\ee
be integrable on $[0,\pi]$. 
Then assumption (\ref{eq-integrability_assumption}) on the correlation function $f_i(x,t)$
is satisfied in dimension $d \geq 2$. 
\end{prop}
%%%%%%%%%%%%%%%%%%%%%%%%%%%%%%%%%%%%%%%%%%%%%%%%%%%%%%%%%%%%%%%%%

If the bosons are initially at thermal equilibrium, in such a way that
$\zeta_i (q) = ( \e^{\beta |q|}-1)^{-1}$, then the assumptions on $\psi_{i,\pm}$
in Proposition~\ref{prop-decay_correl_funct} 
are satisfied if $g_{0,i} (|q|) \in C^2 (]0,\pi])$ and  
\be
|q|^{\min\{d-4,\frac{d-3}{2} \} } |g_{0,i} |^2
\quad, \quad |q|^{d-3} \frac{\D}{\D |q|} |g_{0,i} |^2
\quad \text{, and} \quad  
|q|^{d-2} \frac{\D^2}{\D |q|^2}  |g_{0,i} |^2
\ee
are  integrable on $[0,\pi]$.

\begin{rem}
For most natural models, 
assumption (\ref{eq-integrability_assumption}) fails in dimension $d=1$;  
see Appendix~\ref{app-C} for an explicit computation. 
\end{rem}

\begin{rem}
For physical applications it would be of interest to estimate the
error terms in the convergence to the scaling limit, that is, to have an explicit 
bound on
$\| \rhoP (\lambda^{-2} \tau)- \e^{-\I \lambda^{-2} \tau [S,\cdot]}\rhosl (\tau)  \|_1$. 
Due  to the repeated use of the dominated convergence theorem, 
our proof of 
Theorem~\ref{thm: main} does not exhibit such a bound.   
One 
could in principle obtain  the error terms  by assuming some explicit  decay of 
the correlation functions $f_i(x,t)$
as in \cite{deroeckkupiainenphotonbound} (see Appendix~B in this reference).
\end{rem}

\begin{rem}
The choice
of the scaling exponent $\alpha$ in the factor $\lambda^\alpha$ in front of the hopping Hamiltonian $\Hkin$
in (\ref{eq-H_P_bis}) is more dictated by mathematical than by
physical motivations.  
For $\alpha  > 2$ the hopping Hamiltonian $\Hkin$ is absent in the dynamics in the scaling limit and
one has a trivial coupling between 
the hopping and the internal degrees of freedom in the Lindbladian $\Ll^\natural$.
For $\alpha<2$, as explained in the introduction, the convergence in the weak coupling limit $\lambda\rightarrow 0$
is mathematically more involved. It is not clear whether in this case one can still distill a limiting Lindblad operator.
\end{rem}

As already indicated in the introduction, 
in the case $\alpha =2$
 our scaling limit incorporates both features
 of the singular and weak coupling limits: 
the translational degrees of freedom are
treated within the singular coupling limit and the internal 
degree of freedom is treated within the weak coupling limit. 
The fact that the Hamiltonian  
\be \label{eq: ham for discussion}
H_\lambda = \lambda^2 \Hkin  + H_B + \lambda \Hint
\ee
corresponds  in the limit $\la\to0$ to the singular coupling limit  was pointed out by Palmer~\cite{Palmer77}:  
the dynamics generated by \eqref{eq: ham for discussion}
at time $t$  
can be mapped into the dynamics generated by 
the Hamiltonian
\be \label{eq: ham for discussion_bis}
H_\lambda' = \Hkin + H_B ' + {H_{\text{int}}^\lambda }' \ee
at the (unrescaled) time $\tau$ and with an (unrescaled) hopping term $\Hkin$,
where $H_B'$ is a free boson Hamiltonian as in (\ref{eq-free_boson_Hamiltonian}).
The new   interaction Hamiltonian 
${H_{\text{int}}^\lambda }'$ is not multiplied by $\lambda$ and 
is given by the original interaction Hamiltonian  $\Hint$, as given by (\ref{eq-Hint}), 
but with a rescaled form factor
 $g_{\lambda^2 x} (\lambda^2 q)$ instead of $g_x (q)$.
The two dynamics generated by (\ref{eq: ham for discussion}) and (\ref{eq: ham for discussion_bis}) 
are exactly the same, in the sense that
$\e^{\I t H_\lambda}$ and  $\e^{\I \tau H_\lambda '}$ coincides up to a conjugation by 
the unitary operator transforming the field operator $\Phi (\varphi)$ of a
 boson
with wavefunction $\varphi \in L^2 (\torus^d)$ in the initial problem onto 
the field operator $\Phi ( \varphi_\lambda)$ of a boson in the new problem, with 
$\varphi_\lambda (q) = \lambda \varphi ( \lambda^2 q )$.  
In the limit $\lambda \rightarrow 0$, the rescaled form factor becomes singular.

Let us mention that a model 
of a quantum system coupled to a free fermion bath has been studied in~\cite{Hellmich}
 in the singular
coupling limit, using the approach of Refs.~\cite{Davies74,Davies76}.

%%%%%%%%%%%%%%%%%%%%%%%%%%%%%%%%%%%%%%%%%%%%%%%%%%%%%%%%%%%%%%%%%%%%%%%%%%%
\subsection{A jump process in momentum space} \label{sec: setup equilibrium}
%%%%%%%%%%%%%%%%%%%%%%%%%%%%%%%%%%%%%%%%%%%%%%%%%%%%%%%%%%%%%%%%%%%%%%%%%%%%

Let us choose
$H_{\rm{hop}}= - \Delta$ according to (\ref{eq-discrete_Laplacian}), $\alpha=2$, and 
assume that all baths are initially in thermal equilibrium with the same temperature
$\beta_i^{-1}= \beta^{-1}$.
We may then just as well consider only one bath initially in the Gibbs state
$\rho_B^\La = \rho_\beta^\La$.

\begin{figure}[ht!] 
\vspace{0.5cm}
  \centering
\psfrag{inmomentum}{ $k$}
\psfrag{outmomentum}{ $k'$}
\psfrag{emittedmomentum}{ $q$}
\psfrag{absorbedmomentum}{ $q$}
\psfrag{inlevel}{ $E_s$}
\psfrag{outlevel}{ $E_{s'}$}
  \subfloat[The particle makes a transition  $(k,s) \rightarrow (k',s')$ and
emits a boson of momentum 
$q$ and energy $\nu(q)$ with
$k=k'+q$ and $E_{s}=\nu(q)+E_{s'}$.] 
{\label{fig: markov 1} 
\includegraphics[width = 7cm, height=3.5cm]{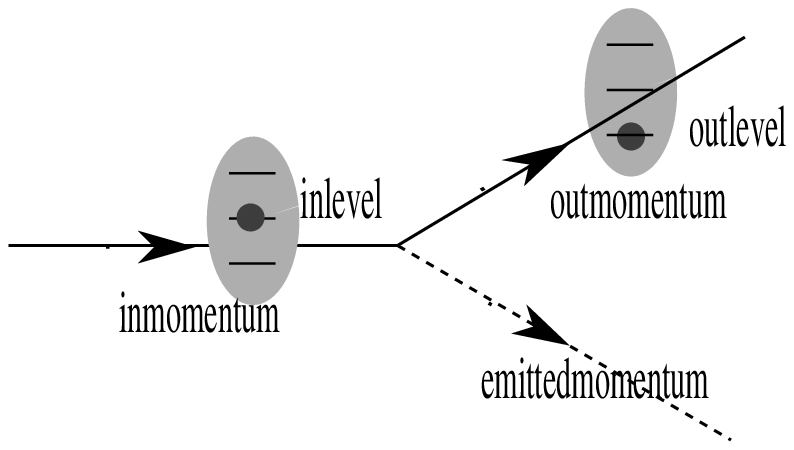}     } \hspace{0.5cm}
  \subfloat[The particle makes a transition  $(k,s) \rightarrow (k',s')$ and 
absorbs a boson of momentum $q$ with 
$k+q=k'$ and $E_s+\nu(q)= E_{s'}$.]   
{\label{fig: markov 2} \includegraphics[width = 7cm, height=3.5cm]{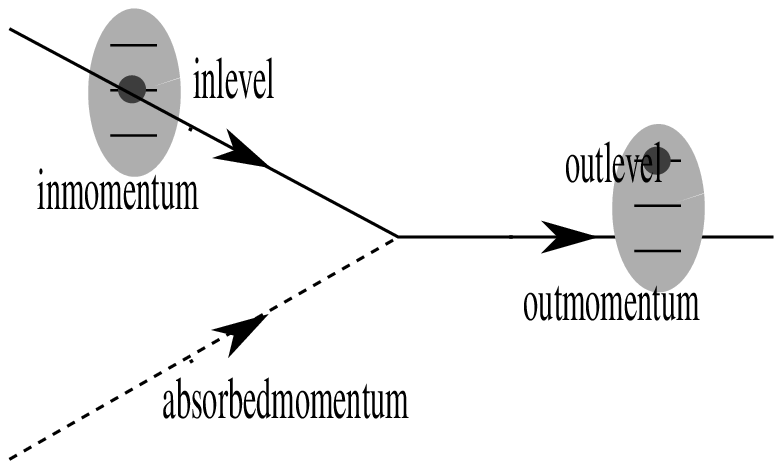}    }
\caption{ \footnotesize{The processes contributing to the gain term 
(first term on the right-hand side of \eqref{eq: explicit caM00}). Emission corresponds to 
$E_s>E_{s'}$ and absorption to $E_s<E_{s'}$. }  }
\label{fig: markov}
\end{figure}

This setup is  interesting if one sets out to study diffusion and decoherence 
of the Brownian particle, see \cite{deroeckdiffusionhighdimension}.
The resulting master equation can be written as
 \begin{eqnarray} \label{eq-Llbis  equilibrium}
\nonumber
\frac{\D\rhosl (\tau)}{\D \tau}   
& = &
   - \I \bigl[- \Delta + \Upsilon , \rhosl (\tau) \bigr]
    + \sum_{\omega \in \sigma( [S,\cdot])} \int_{\torus^d} 
     \frac{\D q}{(2 \pi)^d} \, \widehat{c}(q,\omega)\, 
\biggl( 
 W_\omega\otimes T_q\, \rhosl (\tau) \, W_\omega^\ast\otimes T_q^\ast
\\
& &  
-\frac{1}{2} \Bigl\{ W^\ast_\omega W_\omega\otimes 1_{\ell^2 (\Zd)}\,,\, \rhosl (\tau) \Bigr\}   
\biggr)
\end{eqnarray}
where $T_q = \e^{\I q \cdot X}$ is the unitary momentum translation operator on 
$\ell^{2}(\Zd)$ and we have used the notation 
$\widehat{c}_i (\D q, \omega) = \widehat{c}(q,\omega) \D q $
with $ \widehat{c}(q,\omega)$ the positive distribution which for $\omega \neq 0$ is given by
(see (\ref{eq-assumption}), (\ref{eq: expression c}), and (\ref{eq-FT_c})) 
 \be \label{eq-forumla_c(q,omega)}
\widehat{c} (q,\omega) = 
 2 \pi  \str g_{0} (q)\str^2  \Bigl( 
\zeta(-q)   \delta(\nu(-q)-\omega)  + (1+\zeta(q))    \delta(\nu(q) +\omega) \Bigr) \;.
 \ee
Here $\zeta(q)= (\e^{\beta \nu(q)}-1)^{-1}$ is the Bose-Einstein distribution
and we have assumed that  $\widehat{c}_i (\D q, \omega)$ defines a bona fide measure, 
which requires some mild additional conditions on $\nu$.  
The first term in the rate (\ref{eq-forumla_c(q,omega)}) corresponds to  absorption processes of a boson
with momentum $-q$ and frequency $\nu (-q)=\omega$; it is proportional to the mean number 
$\zeta (-q)$ of bosons with momentum $-q$. The second term corresponds to 
spontaneous and stimulated emission of a boson with momentum $q$ and
frequency $\nu(q)=-\omega$ and is proportional to 
$1+\zeta(q)$ (the term $1$ comes from spontaneous emission).
The delta distributions in (\ref{eq-forumla_c(q,omega)}) accounts for energy conservation
(see Figure~\ref{fig: markov}). 
Note that we have the detailed balance condition 
$ \widehat{c}(q,\omega)   = \e^{-\beta \omega}  \widehat{c}(-q,-\omega)$.
 
To appreciate the equation \eqref{eq-Llbis equilibrium}, it is worthwhile to see what 
it implies for the evolution of diagonal elements of the density matrix in the eigenbasis 
of $S$ and the momentum basis for the translational degrees of freedom. Let us define 
(assuming that the sum on the right-hand side is absolutely convergent)
\be \label{eq-momentum_density}
\rhosl (\tau ; k,s) = \sum_{x,x'\in \Zd}  \e^{\I (x-x') \cdot k}   \bra{x, s} 
\rhosl (\tau) \ket{ x', s}
\ee 
with $\ket{x,s} = \ket{x} \otimes \ket{s}$.
The momentum density (\ref{eq-momentum_density}) satisfies
$\sum_{s=1}^N \int_{\torus^d} \D k\,\rhosl (\tau; k,s)/(2\pi)^d = 1$ for any $\tau\geq0$.
For simplicity, let us assume that the spectrum of $S$ is non-degenerate, \ie, 
all eigenvalues of $S$ are simple.
Then \eqref{eq-Llbis equilibrium} gives  
\be \label{eq: explicit caM00}
\frac{\partial}{\partial \tau} \rhosl (\tau; k',s')  
 =   
  \sum_{s=1 }^N \int_{\torus^d} \D k     
   \Bigl(   \gamma (k',s'| k,s) \, \rhosl (\tau;k,s)  
     -  \gamma (k,s |  k',s') \, \rhosl (\tau;k',s')  \Bigr)
\ee
where 
\be \label{eq-transition_rates} 
\gamma (k',s'| k,s)
 = 
  (2\pi)^{-d} \,\widehat{c}(k-k',E_{s'}-E_s) \str \bra{s'} W \ket{s} \str^2 \;.
\ee 
In formula \eqref{eq: explicit caM00} one recognizes the structure of a forward Markov generator
with (singular) transition rates $\gamma (k',s'| k,s)$, 
acting on  densities of absolutely continuous probability measures (hence on $L^1$-functions) on 
$\torus^d \times \{ 1,\cdots, N\}$.
Therefore, the master equation (\ref{eq-Llbis  equilibrium})
describes the stochastic evolution of a particle with momentum $k$ and internal state
 $s$, which may jump from the state $(k,s)$ to $(k',s')$ by emitting or absorbing 
a boson of momentum $q$ and energy $\nu(q)$, as represented in Figure~\ref{fig: markov}.
According to (\ref{eq-forumla_c(q,omega)}) and 
(\ref{eq-transition_rates}),
 only jumps satisfying 
energy and momentum conservation  $\nu (q) = |E_{s'} - E_s|$ and $q={\rm{sign}}(E_{s'}-E_s) (k'-k)$ are allowed
(here $\rm{sign}$ is the sign function).
Note that,  in the limit 
$\lambda\rightarrow 0$, the energy of the particle coincides with its 
internal energy $E_s$ since the   hopping energy was assumed to be of the order of 
$\lambda^\alpha$ with $\alpha\geq 2$. 
This explains why the detailed balance condition
\be \label{eq: detailed balance}
\gamma (k',s' | k,s)   
 = \e^{\beta(E_s-E_{s'})} \gamma (k,s | k' ,s' ) 
\ee
does not involve  the  kinetic  energy but only the internal energy 
levels $E_s$ and $E_{s'}$.

Let us recast the master equation (\ref{eq: explicit caM00}) in a more explicit form, 
making some concrete choices. 
We assume that $N=2$, label the two spin states as $s\in \{ -,+\}$, and choose the two
internal energies $E_{\pm}=\pm \epsilon/2$. Furthermore, we suppose 
that on $\{\str q \str \leq \epsilon\}$,  the form factor $g_0(q)$ depends only on $|q|$ 
and one has a linear dispersion relation $\nu(q)=\str q \str$. 
Then 
\be \label{eq: more explicit caM00}
\frac{\partial}{\partial \tau} \rhosl (\tau; k',\pm )  
 =   c
 \int_{S^{d-1}(\epsilon)} \D q      \, 
   \Bigl(   \e^{\mp \beta \epsilon/2}  \, \rhosl (\tau;k'-q,\mp )  
     -  \e^{\pm \beta \epsilon/2} \, \rhosl (\tau;k',\pm )  \Bigr)
\ee
where $S^{d-1}(\epsilon)$ is the hypersphere with radius $\epsilon$, $\D q$ is its surface measure, 
and the prefactor $c$ is equal to
 $(2 \pi)^{1-d} | \langle +  | W | - \rangle |^2 |g_0(|q|=\epsilon) |^2 e^{\beta \epsilon/2} /( e^{\beta \epsilon} -1 )$.

In the absence of internal Hamiltonian (\ie, for $S=0$),
energy and momentum conservation and the fact that 
$\nu(q)\not= 0$ for $q \not= 0$ imply that, up to second order in $\lambda$,
the particle can only emit or absorb 
zero-momentum bosons and thus cannot change its momentum.
Hence the coupling with the bath has no effect on the translational degrees of freedom in the scaling limit
(\ref{eq: limit of main result}).
This features would not change if we consider baths at different temperatures, or more complicated hopping 
Hamiltonians. 
 
  Given some mild technical conditions, one can show that a particle described by the Lindblad equation \eqref{eq-Llbis  equilibrium} diffuses in space, i.e.,  
$  \sum_{x \in \mathbb{Z}^d} \str x \str^2 \bra{x}  \rhosl (\tau)  \ket{x} \propto \tau$ as $\tau \to \infty$, 
see \cite{deroeckdiffusionhighdimension, deroeckkupiainendiffusion}.

%%%%%%%%%%%%%%%%%%%%%%%%%%%%%%%%%%%%%%%%%%%%%%%%%%%%%%%%%%%%%%%%%%%%%%%%%%5
%%%%%%%%%%%%%%%%%%%%%%%%%%%%%%%%%%%%%%%%%%%%%%%%%%%%%%%%%%%%%%%%%%%

%%%%%%%%%%%%%%%%%%%%%%%%%%%%%%%%%%%%%%%%%%%%%%%%%%%%%%%%%%%%%
\subsection{Ratchet} \label{sec: setup nonequilibrium}
%%%%%%%%%%%%%%%%%%%%%%%%%%%%%%%%%%%%%%%%%%%%%%%%%%%%%%%%%%%%%%

\begin{figure}[ht] 
\begin{centering}
\vspace{0.5cm}
\psfrag{Pot0}{$\ket{s_{\leftarrow}}$}
\psfrag{Pot1}{$\ket{s_{\rightarrow}}$}
\psfrag{Pot2}{$\ket{s_{\downarrow}}$}
\psfrag{Pot3}{$\ket{s_{\uparrow}}$}
\psfrag{slow}{\hspace*{-3mm} $T_1=0$}
\psfrag{fast}{\hspace*{-3mm} $T_2 >0$}
\includegraphics[width = 8cm, height=6cm]{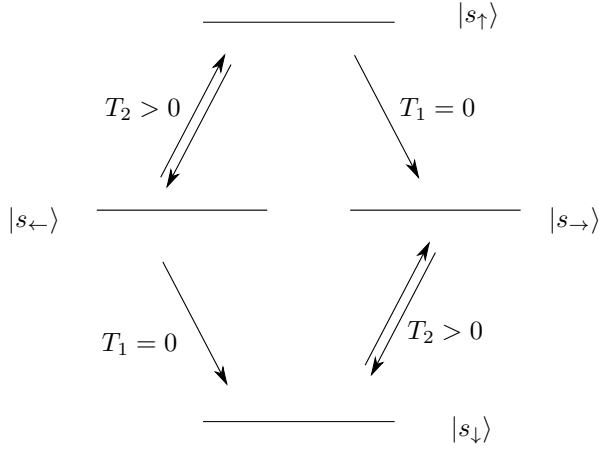} \hspace{0.5cm}
\caption{ \footnotesize{Possible jumps between internal states of the particle. 
 Note that the jumps due to the coupling with the reservoir at temperature
$T_1=0$ can take place only from higher to lower energies, 
whereas the jumps due to coupling with the other
reservoir at positive temperature can go either way.}  }
\label{fig: potentialjumps}
\end{centering}
\end{figure}

We now apply our model to ratchets. We refer the reader to the review article~\cite{Reimann2002}
for more details on this subject. 
For this application, we choose 
\be \label{eq-H_s_x}
\Hkin   =   - \sum_{x \in \Zd}    
\left(   \ketbra{x}{x+e_1} \otimes \ketbra{s_{\rightarrow}}{s_{\leftarrow}}  +    
\ketbra{x+e_1}{x} \otimes \ketbra{s_{\leftarrow}}{s_{\rightarrow}}
\right)\;.
\ee
Here,
$\ket{s_{\rightarrow}}$ and $\ket{s_{\leftarrow}}$ are two distinguished 
eigenstates of $S$ and 
we singled out a spatial direction by coupling these states to the
motion in the direction of the unit vector $e_1 \in \Zd$. 
We assume that 
$S$ has four distinct eigenstates labelled by $s_\uparrow,s_\downarrow, s_\rightarrow$, and 
$s_\leftarrow$ satisfying
\begin{equation}
S \str s_{\uparrow} \rangle =  \epsilon  \str s_{\uparrow} \rangle , \quad  
S \str s_{\downarrow} \rangle =  -\epsilon \str s_{\downarrow} \rangle \quad ,
\quad \text{and} \quad  S \str s_{\rightarrow} \rangle =  S  \str s_{\leftarrow} \rangle =0\;.
\end{equation}
For simplicity, we choose   equal eigenvalues of $S$ corresponding to the states 
$\ket{s_{\rightarrow}}$ and $\ket{s_{\leftarrow}}$, so that
 $\Hkin$ and $S$ commute. 
We will exploit the fact that we have two reservoirs. 
Bosons from the first reservoir couple to the $s$-variable by
\be
W_{i=1} =   \str s_\uparrow  \rangle \langle s_\rightarrow \str +  \str s_\downarrow  \rangle \langle s_\leftarrow \str + \ketbra{s_\rightarrow}{s_\uparrow} 
+ \ketbra{s_\leftarrow}{ s_\downarrow } 
\ee
while bosons of the second reservoir couple as 
\be
W_{i=2} =   \str s_\uparrow  \rangle \langle s_\leftarrow \str +  \str s_\downarrow  \rangle \langle s_\rightarrow \str + \ketbra{ s_\leftarrow}{s_\uparrow} + 
\ketbra{s_\rightarrow}{s_\downarrow}\;. 
\ee
We choose the initial state of the reservoirs to be 
$\rho_B = \rho_{\beta_1}^{(1)} \otimes \rho_{\beta_2}^{(2)}$, where  
$\rho_{\beta_i}^{(i)}$ is a Gibbs (thermal) state at
temperature $T_i = \beta_i^{-1}$.
If the two reservoirs have the same temperature, then the model does not 
display any current. However, by preparing the reservoirs at different temperatures 
$T_1 \not= T_2$
one breaks the time-reversal symmetry and a current will in general emerge. 
To simplify the forthcoming discussion, we choose 
$g_{0,1}=g_{0,2}$ (the form factors are equal) and
\be
T_1= 0,  \qquad   T_2  >0 \;.
\ee
The boson field induces jumps between the internal states $\ket{s}$  as 
represented in Figure~\ref{fig: potentialjumps}. 
By using energy conservation 
(see (\ref{eq-forumla_c(q,omega)})), one easily convinces oneself that, with the temperatures
 chosen as above, the particle can make a transition from $\ket{s_{\leftarrow}}$ 
to $\ket{s_\rightarrow}$ by emitting  a boson of the first reservoir and absorbing one of the second reservoir,
whereas it can not make the reverse transition
from $\ket{s_\rightarrow}$ to $\ket{s_{\leftarrow}}$ (there are no boson with frequency 
$\epsilon$ in the first reservoir at $T_1=0$).
Since all the jumps between eigenstates of $S$ happen at fixed position, 
these transitions do not in themselves induce a current. 
However, the hopping Hamiltonian $\Hkin$  allows for transitions between the states 
$\ket{s_{\rightarrow}}$ and $\ket{s_\leftarrow}$. 
 Hence, a current flows in the $e_1$-direction. 
The possibility of extracting work from the system is already 
visible in Figure~\ref{fig: potentialjumps}, where one sees that the particle can go
from  $\ket{s_{\leftarrow}}$ to $\ket{s_\rightarrow}$  
via the upper (respectively lower) level only clockwise (anticlockwise). 
Once one has this property (which is excluded in equilibrium), it is clear 
that one can devise 
a scheme to convert this ``internal current'' into a spatial current. 

%%%%%%%%%%%%%%%%%%%%%%%%%%%%%%%%%%%%%%%%%%%%%
%%%%%%%%%%%%%%%%%%%%%%%%%%%%%%%%%%%%%%%%%%

%%%%%%%%%%%%%%%%%%%%%%%%%%%%%%%%%%%%%%%%%%%%%%%%%%%%%%%%%%%%%%%%%%%%%%%%%%
\section{Proofs} \label{sec: proof}
%%%%%%%%%%%%%%%%%%%%%%%%%%%%%%%%%%%%%%%%%%%%%%%%%%%%%%%%%%%%%%%%%%%%%%%

%%%%%%%%%%%%%%%%%%%%%%%%%%%%%%%%%%%%%%%%%%%%%%%%%%%%%%%%%%%%
\subsection{Preliminaries}
%%%%%%%%%%%%%%%%%%%%%%%%%%%%%%%%%%%%%%%%%%%%%%%%%%%%%%%%%%%%

Our proof of Theorem~\ref{thm: main} follows a similar approach  as
in some previous works of one of the authors, 
in particular~\cite{deroeckdiffusionhighdimension,deroeck}. The starting point is  
a Dyson expansion of the propagator $\propa{t}$.  
We consider the situation in which the particle is coupled to a single bath.
Therefore, we do not write the lower index $i$ and the corresponding sums, 
which do not play any role and 
obscure the notation. 
The proof for several baths, i.e.,  $\str I \str >1$, is  
exactly the same. 
We also restrict ourselves to the case of a scaling exponent $\alpha=2$. 
The case $\alpha>2$ is simpler and can be treated similarly.

Let us first fix some notation. 
For products of operators, we use the conventions
\be
 \mathop{\prod}\limits_{j=1,\ldots,n}^{\leftarrow} A_j =   A_n \ldots A_2A_1, \qquad    
 \mathop{\prod}\limits_{j=1,\ldots,n}^{\rightarrow} A_j =   A_1 A_2 \ldots  A_n  \;.
\ee
 The operators  acting on $\traceclass$ are denoted by calligraphic fonts, e.g.\ $\mathcal T$ or 
$\mathcal L$, and we use their (operator) norms 
\begin{equation}
\norm \mathcal T \norm =  \sup_{A \in \traceclass} \frac{\norm \mathcal T(A) \norm_1}{\norm A \norm_1}
\; .
\end{equation}

We denote by $C$ constants that can depend only on the space dimension $d$, the internal space dimension $N$, 
the  parameter $\kappa$ in the propagation bound (\ref{eq: propagation estimate}), and the correlation 
function $f(x,t)$.

%%%%%%%%%%%%%%%%%%%%%%%%%%%%%%%%%%%%%%%%%%%%%%%%%%%%%%%%%%%%%%%%%%%%%%%%%%%%%% 
\subsection{The Dyson series} \label{sec-Dyson_expansion}
%%%%%%%%%%%%%%%%%%%%%%%%%%%%%%%%%%%%%%%%%%%%%%%%%%%%%%%%%%%%%%%%%%%%%%%%%%%%

Let $\propaLa{t}$ be the propagator for the reduced dynamics of the particle at finite volume, 
defined by
\be \label{eq-propagator_finite_volume}
\propaLa{t} (\rhoP^\La ) =\tr_B  \Bigl( \e^{-\I t \Htot }   (\rhoP^\La  
\otimes \rho^{\La}_B)\, \e^{\I t \Htot }  \bigr), 
\quad  \quad   \rhoP^\La \in \Ss (\HcP^\Lambda)\;.
\ee
We will show below that 
the Dyson expansion with respect to the  interaction Hamiltonian (\ref{eq-Hint}) 
of this propagator converges in norm; this expansion reads 
\bea \label{eq-Dyson1} 
\propaDysonLa{t}  (\rhoP^\La)  \nonumber
& = &   
\rhoP^\La +  \sum_{n \geq 1} (-\I\lambda)^n 
\int_{0\leq t_1 \leq \cdots \leq t_n\leq t} 
\D t_1 \cdots \D t_n
\sum_{(x_1,\cdots , x_n) \in \La^{n}}
\\ 
&  &
\tr_{B}  \Bigl( 
\Bigl[  V_{x_n}^{\La} (t_n) \otimes \Phi^\La_{x_n} (t_n) \,,\, 
\cdots \Bigl[  V_{x_2}^{\La} (t_2) \otimes \Phi^\La_{x_2} (t_2) \,,\, 
  \bigl[ V_{x_1}^{\La} (t_1) \otimes \Phi^\La_{x_1} (t_1) \, , \,
    \rhoP^\La \otimes \rho_B^\La \bigr] \Bigr] \cdots  \Bigr]   \Bigr)
\eea
where $\Phi^{\La}_{x} (t)= \Phi^\La ( \e^{\I t \nu} g_{x})$ is the freely-evolved  
field operator, see (\ref{eq-evolved_field_op}),
and
\be \label{eq-V_x(t)}
V_{x}^{\La}(t) = \e^{\I t \HPLa }    W \otimes \ketbra{x}{x} \, \e^{-\I t \HPLa }.
\ee
For any $x \in \Zd$, $t \geq 0$, and $T \in \Bb(\HcP^\La)$, let us set 
\be \label{eq-def_cal_I}
\Ii^{\La} (x, t,l) (T) 
 = 
 \begin{cases} 
  V_{x}^{\La} (t)  T 
& \text{ if $l=L$}
\\[2mm]
  - T \,   V_{x}^{\La} (t) 
& \text{ if $l=R$.}
\end{cases}
\qquad , \qquad 
f^\La(x,t,l) = 
\begin{cases}
f^\La(x,t) 
& \text{ if $l=L$}
\\[2mm]
\overline{f^\La (x,t)}
& \text{ if $l=R$}
\end{cases}
\ee
where  $f^\La(x,t)$ is the correlation function (\ref{eq-correl_function}). 
Note that $\overline{f^\La(x,t)} = f^\La (-x, -t)$.
By using the quasi-freeness assumption {\it (B2)}, one gets 
\be \label{eq-Dyson2}
\propaDysonLa{t}
= 1 + \sum_{n\geq 1}\;\;
\sum_{\text{pairings $\underline{\pi}$}}\;\;
\sum_{(\xv,\lv) \in\, \La^{2n}  \times  \{ L, R\}^{2n}   }
 \mathop{\int}\limits_{0 \leq t_1 \leq \cdots\leq t_{2n} \leq t} \D \tv \,
\Vv^{\La}_{\lambda,n} ( \underline{\pi} , \tv, \xv, \lv)
\ee
where  $\D \tv$ stands for $\D t_1 \cdots \D t_{2n}$,  the sum runs over all pairings 
$\underline{\pi} = \{ ( \firstpairing_1,\sigma_1), \cdots, (\firstpairing_n , \sigma_n)\}$
of $( 1, \cdots, 2 n )$, and 
 \be \label{eq-def-Vv} 
\Vv^{\La}_{\lambda,n} ( \underline{\pi} , \tv, \xv, \lv) 
= 
(-\lambda^2)^n  \mathop{\prod}\limits^{\leftarrow}_{j=1,\cdots, 2n} 
\Ii^{\La} (x_{j},t_{j},l_{j}) 
\prod_{m=1}^n f^\Lambda ( x_{\sigma_m} - x_{\firstpairing_m}, t_{\sigma_m} - t_{\firstpairing_m}, l_{\firstpairing_m} )
 \ee
if $x_1,\cdots, x_{2n} \in \Lambda$, and $\Vv^{\La}_{\lambda,n} ( \underline{\pi} , \tv, \xv, \lv)=0$
otherwise. 
We do not write explicitly the dependence of $\HP^\La$, $V_x^\La$, and $\Ii^{\La}$ on the 
coupling constant $\la$ to simplify notation, but we keep it in $\Vv^{\La}_{\lambda,n}$ and $\propaDysonLa{t}$
because we will later consider the limit $\la \rightarrow 0$ of these quantities.

Already at this point, we can establish the norm-convergence of the series \eqref{eq-Dyson2}. Indeed,  
 $\norm \Ii^{\La} (x,t,l) \norm \leq \| W \|$ and  $\str f^\Lambda ( x, t, l ) \str \leq f^{\La}(0,0)$, thus 
the $n$th term in the series (\ref{eq-Dyson2}) has a norm  bounded by 
\be
(2 \str\La\str \lambda \| W\|  )^{2n}  \frac{t^{2n}}{(2n)!} (f^{\La}(0,0))^n    
\left[ 2^{-n}\binom {2n} {n} n! \right]  
= \frac{\bigl(  ( \str\La\str \lambda \| W \| t )^2 2 f^{\La}(0,0) \bigr)^n}{n!}  
\ee
where the term between the square brackets $[\cdot ]$ is the number of pairings $\underline{\pi}$ of
$(1,\cdots,2n)$.  
Hence the Dyson series \eqref{eq-Dyson2} at finite volume converges in norm.
One can prove
that its sum is equal to the propagator in the interaction picture,
\be \label{eq-propa=Dyson_series}
\propaDysonLa{t} =   \e^{\I t [H^{\La}_P, \cdot]} \propaLa{t}\;.
\ee

We will consider the Dyson series at infinite volume and we simply drop the superscript $\La$ on 
$\HP,\Ii,  \Vv$, and  $f$  to denote the corresponding objects for $\La= \Zd$. 
We argue that  in the infinite volume limit $\Lambda \uparrow \Zd$,
\be  \label{eq: three claims}
\e^{-\I t [\HPLa, \cdot] }  \to  \e^{-\I t [\HP, \cdot] } ,  \qquad   \Ii^{\La} (x,t,l) \to  \Ii(x,t,l), \qquad    
\Vv^{\La}_{\la,n} ( \underline{\pi} , \tv, \xv, \lv) \to \Vv_{\la, n} ( \underline{\pi} , \tv, \xv, \lv) 
\ee
strongly on $\traceclass$. Indeed, recall that given some bounded operators $A^\Lambda$ on a Hilbert space
such that  $A^\Lambda \rightarrow 0$ and $(A^\Lambda)^\ast \rightarrow 0$ strongly, then 
$\| A^\Lambda T \|_1 \rightarrow 0$ and 
$\| T A^\Lambda  \|_1 \rightarrow 0$ for any $T \in \traceclass$. 
The first limit in \eqref{eq: three claims} follows from this property, the inequality 
$\| A^\Lambda T U \|_1 \leq \| A^\Lambda T\|_1$ for $U$ unitary, and the strong convergence of 
$\e^{-\I t \HP^\La}$ on $\caH_P$, see \eqref{strong_convergence_unitary_free_ev}. 
As 
 $V_x^\Lambda (t) \rightarrow V_x (t)$ strongly on $\caH_P$ 
(again by \eqref{strong_convergence_unitary_free_ev})), the second limit follows from the same 
property.  
Since $\Vv^{\La}_{\la, n} ( \underline{\pi}, \tv, \xv, \lv)$ is a finite product of the $\Ii$-operators and 
correlation functions $f^{\La}$,  the third limit  in \eqref{eq: three claims} then follows from the second one
and from the pointwise convergence $f^{\La} \to f$ (see Section \ref{sec: thermo limit}).
This implies that term-by-term, the series $\propaDysonLa{t}$ converges strongly to the infinite-volume 
Dyson expansion 
\be \label{eq-Dyson_expansion_therm_limit}
\propaDyson{t} =    \sum_{n} 
\sum_{\underline{\pi}}
\sum_{\xv,\lv   }
 \mathop{\int}\limits_{Z_{2n}(t)} \D \tv  \, 
\Vv_{\lambda,n} ( \underline{\pi} , \tv, \xv, \lv)\;.
\ee
The convergence of this series has not been addressed yet, so for the moment we consider it as a formal series.
To prove Proposition~\ref{lem: thermo limit}, we will show that this convergence is 
in some way uniform in $\Lambda$ and apply the dominated convergence theorem.
We have made in (\ref{eq-Dyson_expansion_therm_limit}) the following abbreviations, which will also be in 
place in the remaining of the paper:
\begin{itemize}
\item  we sum over $n=0,1,\ldots$, where it is understood that the term corresponding to $n=0$ is equal to $1$;
 \item  the sum over $\underline{\pi}$ ranges over all pairings of $(1,2,\cdots ,2n)$; 
\item $Z_n(t)$ denotes the simplex  $\{ \underline{t} = (t_1,\cdots , t_n) \in [0,t]^{n} ;  
t_1 \leq t_2 \leq \cdots \leq t_n \}$;
 \item the sum over $\underline{x}$ and $\underline{l}$ range over $ \mathbb{Z}^{2 n d}$ and 
$ \{\links, \rechts \}^{2n}$, respectively. 
\end{itemize}

%%%%%%%%%%%%%%%%FIGURE 1 %%%%%%%%%%%%%%%%%%%%%%%%%%%%%%%%%%%%%%%%%%%%%
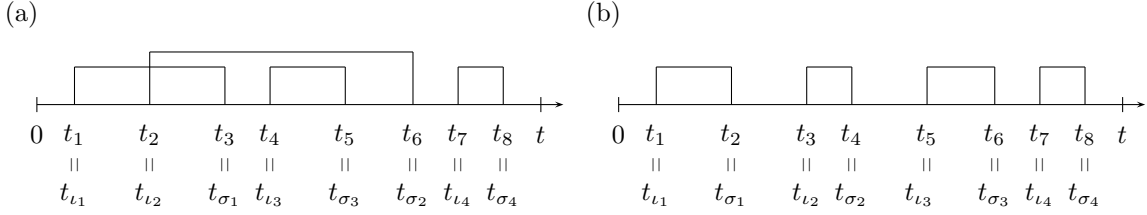
\begin{figure} 
\begin{minipage}{7.2cm}
\begin{pspicture}(-1cm,-1cm)(1cm,1cm)
\psset{linecolor=black}
\psline[linewidth=0.4pt]{->}(7,0)
\rput(-0.2,1.2){(a)}
\psline[linewidth=0.4pt]{}(0,-0.1)(0,0.1)
\rput(0,-0.4){$0$} 
\psline[linewidth=0.4pt]{}(6.7,-0.1)(6.7,0.1)
\rput(6.7,-0.4){$t$} 
\psline[linewidth=0.4pt]{}(0.5,0)(0.5,0.5)
\psline[linewidth=0.4pt]{}(2.5,0)(2.5,0.5)
\psline[linewidth=0.4pt]{}(0.5,0.5)(2.5,0.5)
\rput(0.5,-0.4){$t_1$}
\rput(2.5,-0.4){$t_3$}
\rput(0.5,-0.8){\tiny{$||$}}
\rput(2.5,-0.8){\tiny{$||$}}
\rput(0.5,-1.2){$t_{\firstpairing_1}$}
\rput(2.5,-1.2){$t_{\sigma_1}$}
\psline[linewidth=0.4pt]{}(3.1,0)(3.1,0.5)
\psline[linewidth=0.4pt]{}(4.1,0)(4.1,0.5)
\psline[linewidth=0.4pt]{}(3.1,0.5)(4.1,0.5)
\rput(3.1,-0.4){$t_4$}
\rput(4.1,-0.4){$t_5$}
\rput(3.1,-0.8){\tiny{$||$}}
\rput(4.1,-0.8){\tiny{$||$}}
\rput(3.1,-1.2){$t_{\firstpairing_3}$}
\rput(4.1,-1.2){$t_{\sigma_3}$}
\psline[linewidth=0.4pt]{}(1.5,0)(1.5,0.7)
\psline[linewidth=0.4pt]{}(5,0)(5,0.7)
\psline[linewidth=0.4pt]{}(1.5,0.7)(5,0.7)
\rput(1.5,-0.4){$t_2$}
\rput(5,-0.4){$t_6$}
\rput(1.5,-0.8){\tiny{$||$}}
\rput(5,-0.8){\tiny{$||$}}
\rput(1.5,-1.2){$t_{\firstpairing_2}$}
\rput(5,-1.2){$t_{\sigma_2}$}
\psline[linewidth=0.4pt]{}(5.6,0)(5.6,0.5)
\psline[linewidth=0.4pt]{}(6.2,0)(6.2,0.5)
\psline[linewidth=0.4pt]{}(5.6,0.5)(6.2,0.5)
\rput(5.6,-0.4){$t_7$}
\rput(6.2,-0.4){$t_8$}
\rput(5.6,-0.8){\tiny{$||$}}
\rput(6.2,-0.8){\tiny{$||$}}
\rput(5.6,-1.2){$t_{\firstpairing_4}$}
\rput(6.2,-1.2){$t_{\sigma_4}$}
\end{pspicture}
\end{minipage}
\hspace{0.7cm}
%%%%%
\begin{minipage}{7.2cm}
\begin{pspicture}(-0.6cm,-0.6cm)(0.6cm,0.6cm)
\psset{linecolor=black}
\psline[linewidth=0.4pt]{->}(7,0)
\rput(-0.2,1.2){(b)}
\psline[linewidth=0.4pt]{}(0,-0.1)(0,0.1)
\rput(0,-0.4){$0$} 
\psline[linewidth=0.4pt]{}(6.7,-0.1)(6.7,0.1)
\rput(6.7,-0.4){$t$} 
\psline[linewidth=0.4pt]{}(0.5,0)(0.5,0.5)
\psline[linewidth=0.4pt]{}(1.5,0)(1.5,0.5)
\psline[linewidth=0.4pt]{}(0.5,0.5)(1.5,0.5)
\rput(0.5,-0.4){$t_1$}
\rput(1.5,-0.4){$t_2$}
\rput(0.5,-0.8){\tiny{$||$}}
\rput(1.5,-0.8){\tiny{$||$}}
\rput(0.5,-1.2){$t_{\firstpairing_1}$}
\rput(1.5,-1.2){$t_{\sigma_1}$}
\psline[linewidth=0.4pt]{}(2.5,0)(2.5,0.5)
\psline[linewidth=0.4pt]{}(3.1,0)(3.1,0.5)
\psline[linewidth=0.4pt]{}(2.5,0.5)(3.1,0.5)
\rput(2.5,-0.4){$t_3$}
\rput(3.1,-0.4){$t_4$}
\rput(2.5,-0.8){\tiny{$||$}}
\rput(3.1,-0.8){\tiny{$||$}}
\rput(2.5,-1.2){$t_{\firstpairing_2}$}
\rput(3.1,-1.2){$t_{\sigma_2}$}
\psline[linewidth=0.4pt]{}(4.1,0)(4.1,0.5)
\psline[linewidth=0.4pt]{}(5,0)(5,0.5)
\psline[linewidth=0.4pt]{}(4.1,0.5)(5,0.5)
\rput(4.1,-0.4){$t_5$}
\rput(5,-0.4){$t_6$}
\rput(4.1,-0.8){\tiny{$||$}}
\rput(5,-0.8){\tiny{$||$}}
\rput(4,-1.2){$t_{\firstpairing_3}$}
\rput(5,-1.2){$t_{\sigma_3}$}
\psline[linewidth=0.4pt]{}(5.6,0)(5.6,0.5)
\psline[linewidth=0.4pt]{}(6.2,0)(6.2,0.5)
\psline[linewidth=0.4pt]{}(5.6,0.5)(6.2,0.5)
\rput(5.6,-0.4){$t_7$}
\rput(6.2,-0.4){$t_8$}
\rput(5.6,-0.8){\tiny{$||$}}
\rput(6.2,-0.8){\tiny{$||$}}
\rput(5.6,-1.2){$t_{\firstpairing_4}$}
\rput(6.2,-1.2){$t_{\sigma_4}$}
\end{pspicture}
\end{minipage}
\vspace{0.5cm}
\caption{\label{fig-1}
(a) crossed diagram\;
(b) ladder diagram. 
}
\end{figure}

%%%%%%%%%%%%%%%%%%%%%%%%%%%%%%%%%%%%%%%%%%%%%%%%%%%%%%%%%%%%%%% 
\subsubsection{Graphical representation}
%%%%%%%%%%%%%%%%%%%%%%%%%%%%%%%%%%%%%%%%%%%%%%%%%%%%%%%%%%%%%%%%

It is convenient to represent $\Vv_{\lambda,n}$  or $\Vv^{\La}_{\lambda,n}$ by a diagram
in which all times $t_1 \leq t_2\leq \cdots  \leq t_{2n}$ are ordered on the real line 
and pairings are represented by bridges
linking two distinct times. Two examples of diagrams are represented in Figure~\ref{fig-1}.
Replacing the times by their indices $1,2,\cdots,  2n$, the diagrams with 
$2n$ points  are in one-to-one correspondence with the pairings
of $(1,2,\cdots,2n)$.   
A diagram containing two pairs $(\firstpairing,\sigma)\in \underline{\pi}$ and 
$(\firstpairing',\sigma')\in \underline{\pi}$ such that $\firstpairing< \firstpairing'$ and 
$\firstpairing'<\sigma$ is called 
a {\it crossing diagram}\,\footnote{Traditionally, some of these diagrams are called ``nested'', but we call all of
them ``crossing diagrams'' as we represent them on a single time axis, see Figure~\ref{fig-1}.}. 
A non-crossing diagram
 will be called a {\it ladder diagram}; it corresponds to  the pairing 
$\underline{\pi}_{\rm ladder}=\{ (1,2),(3,4),\cdots, (2n-1,2n)\}$ respecting the order, see 
 Figure~\ref{fig-1}.  

 \vspace{1mm}

The 
strategy of the proof
of Theorem~\ref{thm: main} consists in showing that in the scaling limit 
$\lambda \rightarrow 0$, $t = \lambda^{-2} \tau \rightarrow \infty$,
\begin{itemize}
\item all crossing diagrams in the infinite-volume Dyson series  
(\ref{eq-Dyson_expansion_therm_limit})
converge to zero;  

\item the   
ladder diagram in (\ref{eq-Dyson_expansion_therm_limit}) of order $2n$ converges 
to the corresponding  
diagram of the Dyson expansion of 
$\e^{\I \tau [ \Hkin^\natural, \cdot ]} \e^{\tau  \Ll^\natural}$, where $\Ll^\natural$ is the Lindblad generator given in
(\ref{eq-Lindblad_generator}).
\end{itemize}
%

%%%%%%%%%%%%%%%%%%%%%%%%%%%%%%%%%%%%%%%%%%%%%%%%%%%%%%%%%%%%%%%%%%%%%%%%%%5
\subsection{Plan of the proof}
\label{sec-sketch_proof}
%%%%%%%%%%%%%%%%%%%%%%%%%%%%%%%%%%%%%%%%%%%%%%%%%%%%%%%%%%%%%%%%%%%%%%%%%%%
 
Below,  we  give the main  steps of the proof of Theorem~\ref{thm: main}. 

%%%%%%%%%%%%%%%%%%%%%%%%%%%%%%%%%%%%%%%%%%%%%%%%%%%%%%%%%%%%%%%%%%%%%%%%%%
 \subsubsection{Topology} \label{sec: norms}
%%%%%%%%%%%%%%%%%%%%%%%%%%%%%%%%%%%%%%%%%%%%%%%%%%%%%%%%%%%%%%%%%%%%%%%%%%

We first introduce a notion of convergence that is particularly useful for the problem.
Let $(\Ttt_\lambda)_{\lambda}$ be a family of operators on $\traceclass$. 
We associate to $\Ttt_\lambda$  a kernel with values in 
operators on the vector space $\spinobs$ of $N\times N$ matrices, defined as follows:
\be \label{eq-def_kernel}
(\Ttt_{\lambda})_{x_0 ,y_0 ;x, y} ( M ) 
= \bra{x}  \Ttt_{\lambda} ( M \otimes \ketbra{x_0}{y_0}  )  \ket{y}
\quad , \quad M \in \spinobs
\;.\ee
We write  
$\ptlim_{\lambda\rightarrow 0} \Ttt_\lambda=0$ 
 whenever 
\be
\nonumber
\lim_{\lambda\rightarrow 0} \sum_{x,y\in \Zd} 
\bigl\| (\Ttt_{\lambda})_{x_0,y_0;x, y} \bigr\|
 =0
\qquad 
\text{for any $x_0,y_0 \in \Zd$,}
\ee
where the norm inside the sum is the matrix norm.

%%%%%%%%%%%%%%%%%%%%%%%%%%%%%%%%%%%%%%%%%%%%%%%%%%%%%%%%%%%%%%%%%%%%%%%%%%%%%%%
\begin{lemma} \label{def-lem} 
Let $\Ttt_\lambda$ be uniformly bounded operators
on $\traceclass$. If  
$\ptlim_{\lambda\rightarrow 0} \Ttt_\lambda=0$ then
$\Ttt_\lambda \rightarrow 0$ strongly, that is,
$\lim_{\lambda\rightarrow 0} \| \Ttt_\lambda ( T) \|_1 \rightarrow 0$ for any $T \in \traceclass$.
\end{lemma}
%%%%%%%%%%%%%%%%%%%%%%%%%%%%%%%%%%%%%%%%%%%%%%%%%%%%%%%%%%%%%%%%%%%%%%%%%%%%%%%%%%%%

\vspace{1mm}

\proof
Note first that for any $T \in \traceclass$,
\begin{equation} \label{eq-proof_lemma1_1}
\| T \|_1 
 =   
\sup_{A \in \observable} \frac{| \tr ( T A ) |}{\|A\|}
 \leq  
\sup_{A \in \observable} \sum_{x,y \in \Zd}  
  \tr_{\complex^N} ( | \bra{x}T \ket{y} | )  
   \frac{ \|  \bra{y}A \ket{x} \|}{\|A\|}
\leq N
\sum_{x,y \in \Zd}  
  \| \bra{x}T \ket{y}  \|
\end{equation}
where the last inequality follows because 
 $\tr (|M| ) \leq N \|M\|$ for any finite matrix $M \in \caB(\mathbb C^N)$. 
Let $T $ have finite support in the sense that $\bra{x_0}T \ket{y_0}$ is nonzero only for a finite number of 
$x_0,y_0 \in \Zd$. Then, by \eqref{eq-proof_lemma1_1},   
$\ptlim_{\lambda\rightarrow 0} \Ttt_\lambda=0$ implies $ \| \Ttt_\lambda(T)  \|_1 \to 0$. 
Next, one checks that operators $T$ with finite support are dense in $\traceclass$.
Actually, let 
$P_{\Lambda} = \sum_{x \in \Lambda} \ketbra{x}{x} \otimes 1_{\complex^N}$ 
be the 
finite-rank projector on $\Span \{ \ket{x} ; x \in \Lambda \} \otimes \complex^N$, with
$\Lambda= \mathbb{Z}^d/(2L\mathbb{Z})^d$  as before.
Without loss of generality, we may assume that $T \geq 0$. 
If $\{ \ket{\psi_{j}}\}$ is an orthonormal basis of $\HcP$ diagonalizing 
$T$ and $p_j\geq 0$ are the eigenvalues of $T$, then
\begin{eqnarray*}
\bigl\| P_{\Lambda} T P_{\Lambda} - T \bigr\|_1
& \leq   & \sum_j p_{j}  
 \Bigl( 
  \bigl\| P_{\Lambda} \ketbra{\psi_j}{\psi_j} ( P_{\Lambda}- 1) \bigr\|_1
  + \bigl\| ( P_{\Lambda}-1) \ketbra{\psi_j}{\psi_j} \bigr\|_1
 \Bigr)
\\
& \leq & 
2 \sum_{j} p_j
  \bigl\| ( P_{\Lambda}-1) \ket{\psi_j} \bigr\| 
 \quad \rightarrow \quad 0
\end{eqnarray*}
as $L\rightarrow \infty$ by  
dominated convergence.
Hence, for any $T \in \traceclass$ and $\varepsilon>0$, one can
choose a decomposition $T=T_0+T_1$ such that $T_0$ has finite support and $\| T_1  \|_1 \leq \varepsilon$.  
Then  $\| \Ttt_\lambda ( T) \|_1 \leq \| \Ttt_\lambda ( T_0) \|_1 + C \varepsilon $ by the uniform boundedness of 
$(\Ttt_\lambda)_{\lambda}$.  The claim follows. 
\finpro

\vspace{1mm}

A consequence of lemma~\ref{def-lem} is that in order to prove the strong convergence 
of $\propaDyson{\lambda^{-2} \tau}$ to $
\Dd_{\tau}$ in the scaling limit, 
it is enough to show that
\begin{equation} \label{eq-pt_convergence_scaling_limit} 
 \ptlim_{\lambda\rightarrow 0}  
\left\{ \propaDyson{\la^{-2}\tau} 
-  \Dd_{\tau}   \right\}=0 \;.
\end{equation}
In fact, one has 
$\norm  \propaDyson{t}\norm= 1$ for any $t$
by the following standard argument.
One first notes that   the finite volume propagator $\propaLa{t}$ defined in
(\ref{eq-propagator_finite_volume})
preserves positivity and trace and  hence so does the sum $\propaDysonLa{t}$ of its Dyson expansion 
(related
to $\propaLa{t}$ by (\ref{eq-propa=Dyson_series})) as well as
its strong limit 
$ \propaDyson{t}$ as  $\Lambda \uparrow \Zd$ 
(we assume here that  Proposition~\ref{lem: thermo limit} has been already established).
Then the dual of  $\propaDyson{t}$ under the trace,  $\propaDyson{t}^\ast$, which
acts on $ \observable$,   
preserves positivity and satisfies $\propaDyson{t}^\ast(1)=1$.   
It follows that  
  $\| \propaDyson{t}^\ast \|  =1$ (see e.g.~\cite{Bratteli}, Corollary 3.2.6) and thus
$\| \propaDyson{t} \|= 1$.

%%%%%%%%%%%%%%%%%%%%%%%%%%%%%%%%%%%%%%%%%%%%%%%%%%%%%%%%%%%%%%%%%%%%%%%
 \subsubsection{Assumptions} \label{sec: assumptions}
%%%%%%%%%%%%%%%%%%%%%%%%%%%%%%%%%%%%%%%%%%%%%%%%%%%%%%%%%%%%%%%%%%%%%%
 
 In the remainder of this paper, we always assume {\it (A1-A2)} and {\it (B1-B4)} to be valid 
 without further mentioning it 
(note that once the Dyson series for $\propa{t}$ is accepted as the basic object of study, one does  
no need those assumptions anymore). To prove Theorem~\ref{thm: main},  we rely on:
\begin{itemize}
\item[a)] The propagation bound (\ref{eq: propagation estimate}), but for $\La=\bbZ^d$: by 
(\ref{strong_convergence_unitary_free_ev}) and as the 
right-hand side of (\ref{eq: propagation estimate}) is independent of $\La$,  
 this bound remains valid for $\La=\bbZ^d$ (alternatively, one can check this directly by the Combes-Thomas 
estimate, using the fact that $\Hkin$ has a finite range).  
\item[b)]  Assumption (\ref{eq-integrability_assumption}) on the infinite volume correlation function 
$f(x,t)$ or, in most intermediate steps, the weaker requirement that $f(x,\cdot)$ is integrable for any 
$x \in \Zd$.
\end{itemize}

%%%%%%%%%%%%%%%%%%%%%%%%%%%%%%%%%%%%%%%%%%%%%%%%%%%%%%%%%%%%%%%%%%
\subsubsection{Step I}
%%%%%%%%%%%%%%%%%%%%%%%%%%%%%%%%%%%%%%%%%%%%%%%%%%%%%%%%%%%%%

We first prove that 
the Dyson series $\propaDyson{\lambda^{-2} \tau}$, considered as a series in $n$, $\underline{x}$, 
and $\underline{l}$, converges absolutely and uniformly in $\la$, in the sense that 
\be   \label{eq: apriori}
 \sum_n  \sum_{\xv,\lv} 
\sum_{\underline{\pi}}
\sum_{x,y} \sup_{\lambda >0}
\mathop{\int}\limits_{Z_{2n}( \lambda^{-2} \tau)} \D \tv\,
  \bigl\| 
\bigl( \Vv_{\lambda,n} (\underline{\pi},\tv,\xv,\lv) \bigr)_{x_0,y_0;x,y}  
\bigr\|   < \infty
\ee
for any $x_0,y_0 \in \Zd$ (see  Proposition~\ref{lem: sum of dyson} below).
This bound is the crucial point in the proof, since it allows us to estimate the perturbation series term by term.  It relies heavily on the assumption (\ref{eq-integrability_assumption}). 

By a similar bound on the finite-volume Dyson series $\propaDysonLa{t}$,  the fact that 
this series  converges term by term as $\La \uparrow \Zd$, Lemma~\ref{def-lem}, and (\ref{eq: three claims}),
we obtain Proposition~\ref{lem: thermo limit} with 
\be \label{eq-link_propagator_and_Dyson_series}
\propa{t} =   \e^{-\I t [\HP,\cdot] } \propaDyson{t}\;.
\ee
Here the integrability of $f^\La(x,\cdot )$ is not needed, 
the only requirement is its pointwise convergence as $\La \uparrow \Zd$ (which follows from {\it (B4)}).
 
These results are accomplished in subsections \ref{sec: estimating each term} and \ref{sec-thermo_limit}.

%%%%%%%%%%%%%%%%%%%%%%%%%%%%%%%%%%%%%%%%%%%%%%%%%%%%%%%
\subsubsection{Step II}
%%%%%%%%%%%%%%%%%%%%%%%%%%%%%%%%%%%%%%%%%%%%%%%%%%%%%%%

We show that every single crossing diagram vanishes in the scaling limit, in the sense that, for any 
$n \geq 1$ and $x_0,y_0,x,y \in \Zd$,
\be
\lim_{\la \to 0} \mathop{\int}\limits_{Z_{2n}( \lambda^{-2} \tau)} \D \tv  \,
\norm \Vv_{\lambda,n} ( \underline{\pi} , \tv,\xv,\lv )_{x_0,y_0;x,y} \norm =0 \qquad  \text{whenever} \;\; \underline{\pi}\neq  \underline{\pi}_{\rm{ladder}}\;.
\ee

This step is essentially taken over from the original work \cite{Davies74}, but we review it in subsection~\ref{sec-vanishing_crossed_diagrams}. 
To feel why this holds true, note that when $t = \lambda^{-2} \tau \gg 1$,   the time 
integration domain of a crossing diagram of order $2n$ 
is much smaller than that of the ladder diagram of the same order. 
This is due to the restriction 
$t_\firstpairing \leq t_{\firstpairing'} \leq t_{\sigma}$ associated to the nested pairs
$(\firstpairing,\sigma)$ and $(\firstpairing',\sigma')$, see Figure~\ref{fig-1}.

By dominated convergence and \eqref{eq: apriori}, this implies that 
the contribution of crossing diagrams in the Dyson series $\propaDyson{\lambda^{-2}t}$
vanishes in the limit $\la \rightarrow 0$ in the
topology introduced in section~\ref{sec: norms}, \ie,
\be   \label{eq: vaninshing combined}
\sum_n  \sum_{\xv,\lv}  \sum_{x,y } 
\sum_{\underline{\pi} \neq \underline{\pi}_{\rm{ladder}}}
\mathop{\int}\limits_{Z_{2n}( \lambda^{-2} \tau)} \D \tv
 \bigl\| 
\bigl( \Vv_{\lambda,n} (\underline{\pi},\tv,\xv,\lv) \bigr)_{x_0,y_0;x,y}  
\bigr\|   \mathop{\to}\limits_{\la \to 0} 0
\ee
for any $x_0,y_0 \in \Zd$.

%%%%%%%%%%%%%%%%%%%%%%%%%%%%%%%%%%%%%%%%%%%%%%%%%%%%%%%%%%%%%%%%
\subsubsection{Step III} \label{sec: step three}
%%%%%%%%%%%%%%%%%%%%%%%%%%%%%%%%%%%%%%%%%%%%%%%%%%%%%%%%%%%%%%%%

It remains to evaluate the contribution of ladder diagrams $\underline{\pi}_{\rm{ladder}}$. Let 
 $\caK(\underline{\tau},\xv,\lv)$ be the bounded  operators, defined in
Proposition~\ref{prop_spectral_averaging} below, 
which yield  the limiting QDS $(\e^{\tau \caL^\natural})_{\tau \geq 0}$
upon summing over $n$, $\xv$, $\lv$, and $\underline{\tau}$:
\be \label{eq: agreement}
\sum_{n} \mathop{\int}\limits_{Z_n(\tau)}  \D \underline{\tau}\,  \sum_{\xv,\lv} 
\caK_n(\underline{\tau},\xv,\lv) = \e^{\tau \caL^\natural}
\ee
where the sums and integrals are absolutely convergent in norm.
We will show in sections~\ref{sec-ladder_diagrams} and \ref{sec: spectral averaging} 
that 
\be \label{eq: fate of ladders}
\lim_{\la \to 0}\mathop{\int}\limits_{Z_{2n}( \lambda^{-2} \tau)} \D \tv \,  
\Vv_{\lambda,n} (\underline{\pi}_{\rm{ladder}},\tv,\xv,\lv)   
=      \mathop{\int}\limits_{Z_{n}(  \tau)} \D \underline{\tau}\,  
\caK_n (\underline{\tau},\xv,\lv)
\ee
in norm on $\Bb (\traceclass)$,
which of course implies 
\be \label{eq: fate of ladders_bis}
\mathop{\int}\limits_{Z_{2n}( \lambda^{-2} \tau)} \D \tv \,  
\bigl( \Vv_{\lambda,n} (\underline{\pi}_{\rm{ladder}},\tv,\xv,\lv) \bigr)_{x_0,y_0;x,y}
\quad \rightarrow 
 \mathop{\int} \limits_{Z_{2n}(\tau)} \D \tauv \,\bigl( \caK_n (\underline{\tau},\xv,\lv) \bigr)_{x_0,y_0;x,y}\;.
\ee
Dominated convergence allows us to conclude from   (\ref{eq: apriori})
and (\ref{eq: fate of ladders_bis}) that (\ref{eq-pt_convergence_scaling_limit}) holds true.

%%%%%%%%%%%%%%FIGURE 2 %%%%%%%%%%%%%%%%%%%%%%%%%%%%%%%%%%%%%%%%%%%%%
\begin{figure}
\begin{pspicture}(-1cm,-1cm)(1cm,1cm)
%%% one line diagram
\psset{linecolor=black}
\psline[linewidth=0.4pt]{->}(7,0)
\psline[linewidth=0.4pt]{}(0,-0.1)(0,0.1)
\rput(0,-0.3){$0$} 
\psline[linewidth=0.4pt]{}(6.7,-0.1)(6.7,0.1)
\rput(6.7,-0.3){$t$} 
\psline[linewidth=0.4pt]{}(0.8,0)(0.8,0.5)
\psline[linewidth=0.4pt]{}(2.3,0)(2.3,0.5)
\psline[linewidth=0.4pt]{}(0.8,0.5)(2.3,0.5)
\rput(0.8,-0.3){$t_1$}
\rput(2.3,-0.3){$t_3$}
\rput(0.8,-0.8){$x_1$}
\rput(2.3,-0.8){$x_3$}
\psline[linewidth=0.4pt]{}(3.1,0)(3.1,0.5)
\psline[linewidth=0.4pt]{}(4.1,0)(4.1,0.5)
\psline[linewidth=0.4pt]{}(3.1,0.5)(4.1,0.5)
\rput(3.1,-0.3){$t_4$}
\rput(4.1,-0.3){$t_5$}
\rput(3.1,-0.8){$x_4$}
\rput(4.1,-0.8){$x_5$}
\psline[linewidth=0.4pt]{}(1.5,0)(1.5,0.7)
\psline[linewidth=0.4pt]{}(5,0)(5,0.7)
\psline[linewidth=0.4pt]{}(1.5,0.7)(5,0.7)
\rput(1.5,-0.3){$t_2$}
\rput(5,-0.3){$t_6$}
\rput(1.5,-0.8){$x_2$}
\rput(5,-0.8){$x_3$}
\psline[linewidth=0.4pt]{}(5.6,0)(5.6,0.5)
\psline[linewidth=0.4pt]{}(6.2,0)(6.2,0.5)
\psline[linewidth=0.4pt]{}(5.6,0.5)(6.2,0.5)
\rput(5.6,-0.3){$t_7$}
\rput(6.2,-0.3){$t_8$}
\rput(5.6,-0.8){$x_7$}
\rput(6.2,-0.8){$x_8$}
%%%%% two-line diagram
\psline[linewidth=0.4pt]{->}(8,1)(15,1)
\rput(15.5,1){L}
\psline[linewidth=0.4pt]{}(8,0.9)(8,1.1)
\rput(8,0.7){$0$} 
\psline[linewidth=0.4pt]{}(14.7,0.9)(14.7,1.1)
\rput(14.7,0.7){$t$} 
\psline[linewidth=0.4pt]{->}(8,-1)(15,-1)
\rput(15.5,-1){R}
\psline[linewidth=0.4pt]{}(8,-0.9)(8,-1.1)
\rput(8,-1.3){$0$} 
\psline[linewidth=0.4pt]{}(14.7,-0.9)(14.7,-1.1)
\rput(14.7,-1.3){$t$}
\psline[linewidth=0.4pt]{}(8.8,1)(8.8,1.5)
\psline[linewidth=0.4pt]{}(10.3,1)(10.3,1.5)
\psline[linewidth=0.4pt]{}(8.8,1.5)(10.3,1.5)
\rput(8.8,0.7){$t_1$}
\rput(10.3,0.7){$t_3$}
\rput(8.8,1.8){$\tilde{x}_1$}
\rput(10.3,1.8){$\tilde{x}_2$}
%\rput(8.5,0.3){\tiny{$||$}}
%\rput(10.5,0.3){\tiny{$||$}}
%\rput(8.5,0){$t_{i_1}$}
%\rput(10.5,0){$t_{i_2}$}
\psline[linewidth=0.4pt]{}(11.1,1)(12.1,-1)
\rput(11.1,1.3){$t_4$}
\rput(12.1,-1.3){$t_5$}
\rput(11.1,1.8){$\tilde{x}_3$}
\rput(12.1,-1.8){$\tilde{y}_2$}
\psline[linewidth=0.4pt]{}(9.5,-1)(13,1)
\rput(9.5,-1.3){$t_2$}
\rput(13,1.3){$t_6$}
\rput(9.5,-1.8){$\tilde{y}_1$}
\rput(13,1.8){$\tilde{x}_4$}
\psline[linewidth=0.4pt]{}(13.6,-1)(14.3,1)
\rput(13.6,-1.3){$t_7$}
\rput(14.3,1.3){$t_8$}
\rput(13.6,-1.8){$\tilde{y}_3$}
\rput(14.2,1.8){$\tilde{x}_5$}
\end{pspicture}
\vspace{1cm}
\caption{ \label{fig-2}
Equivalent representations of a $n=8$-points diagram: on the right diagram, the times
$t_j$ with $l_j=L$ are put on the upper axis and 
the times $t_k$ with $l_k=R$ on the lower axis.  Here $|l|=5$, $(j_1,j_2,j_3,j_4,j_5)=(1,3,4,6,8)$,
and $(k_1,k_2,k_3)=(2,5,7)$. 
}
\end{figure}
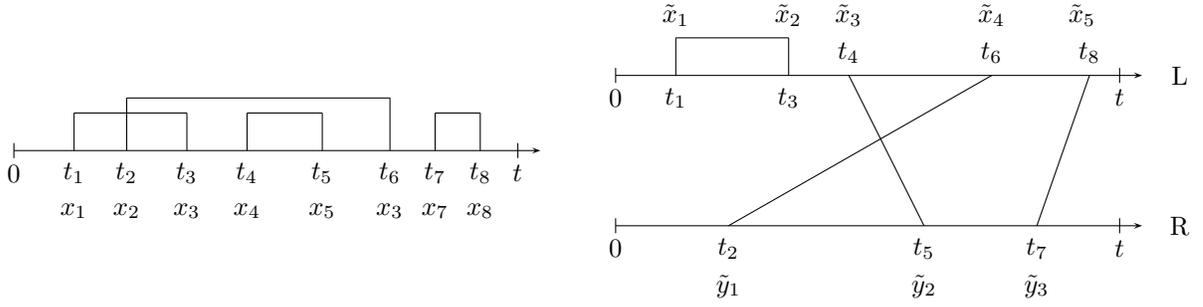

%

%%%%%%%%%%%%%%%%%%%%%%%%%%%%%%%%%%%%%%%%%%%%%%%%%%%%%%%
\subsection{Estimating each term of the Dyson series} \label{sec: estimating each term}
%%%%%%%%%%%%%%%%%%%%%%%%%%%%%%%%%%%%%%%%%%%%%%%%%%%%%%

%%%%%%%%%%%%%%%%%%%%%%%%%%%%%%%%%%%%%%%%%%%%%%%%%%%%%%%%%%%%%%%%%%%%%%%%%%%%%%%%%%%
\begin{lemma} \label{lemma1} Fix $n, \underline{\pi}, x_0,y_0 $,  and $\tv \in Z_{2n}(t)$. Then
\be \label{eq-lemma1}
 \bigl\|  \bigl( \Vv_{\lambda,n} ( \underline{\pi} , \tv,\xv,\lv )\bigr)_{x_0,y_0;x,y} \bigr\| 
\leq  (C \lambda)^{2n} \, \e^{4 \lambda^2 t}\,  
  \prod_{m=1}^n h_{n} (t_{\sigma_m}-t_{\firstpairing_m} )  \,R_{x_0,y_0;x,y}^{(n)}(\xv,\lv )
\ee
where 
\be \label{eq-h_n}
 h_n  = \sup_{x \in \Zd} | f (x,\cdot) | \e^{-\frac{| x|}{2n} }
\ee
and $R_{x_0,y_0;x,y}^{(n)}( \xv,\lv)$ is independent of  $\lambda$ and $\Lambda$ and such that 
\be \label{eq-eq-C_x_0_y_0}
\sup_{n \geq 0} \Bigl\{ \sum_{\xv ,\lv}\, \sum_{x,y} R_{x_0,y_0;x,y}^{(n)} ( \xv,\lv) \Bigr\}  
\;\;< \;\; \infty\;. 
\ee
The bound (\ref{eq-lemma1}) is also true if one replaces 
$\Vv_n$ by $\Vv^{\La}_n$
and $h_n$ by $\max_{x \in \La} | f^\La (x,\cdot ) |$.
\end{lemma}
%%%%%%%%%%%%%%%%%%%%%%%%%%%%%%%%%%%%%%%%%%%%%%%%%%%%%%%%%%%%%%%%%%%%%%%%%%%%%%%%%%%%%

\vspace{4mm}

\proof
We first give an explicit formula for 
$\Vv_{\la,n}(\underline{\pi},\underline{t},\underline{x},\underline{l}) 
( M \otimes \ketbra{x_0}{y_0})$ in terms of the operators $V_x (t)$,
with $M \in \spinobs$ and $x_0 , y_0 \in \Zd$.
For a fixed $\lv \in \{ L , R \}^{2n}$, let
$|l|$ denotes the number of indices $j$ such that 
$l_j =L$, $j=1,\cdots, 2 n$. We change the labelling of the indices and 
coordinates by defining (see Figure~\ref{fig-2}) 
\be \nonumber
\begin{array}{lcl}
\bigl\{ j_1 < \cdots < j_{|l|} \bigr\} 
& = & \bigl\{ j \in \{ 1, \cdots , 2 n \}\; ;\; l_j =L\bigr\}
\\[3mm] 
\{ k_1 < \cdots < k_{2n-|l|} \} 
& = & \bigl\{ k \in \{ 1, \cdots , 2 n \}\; ;\; l_k =R \bigr\}
\end{array}
\ee
and
\be
\begin{array}{lclclclclclcl}
\nonumber
 \tilde{x}_0 & = & x_0
& , & \tilde{x}_1 & = & x_{j_1} & , &  
\cdots & , & \tilde{x}_{|l|} & = & x_{j_{|l|}} 
\\ 
\tilde{y}_0 & = & y_0 & , &  
\tilde{y}_1 & = &  x_{k_1} & , &  \cdots & , & \tilde{y}_{2n-|l|}  & = &  x_{k_{2n-|l|}}
\end{array}
\;.
\ee
It follows from (\ref{eq-def_cal_I}) and (\ref{eq-def-Vv}) 
that for any matrix $M \in \spinobs$, 
\bea \label{eq-Vv_n}
\nonumber
 \Vv_{\la,n} ( \underline{\pi} , \tv, \xv, \lv)
\bigl(M \otimes \ketbra{x_0}{y_0} \bigr) 
&  = &  
\lambda^{2n} (-1)^{n+|l|}
V_{\tilde{x}_{|l|}} (t_{j_{|l|}}) \cdots V_{\tilde{x}_{1}} (t_{j_{1}})
M \otimes \ketbra{ {x}_0}{ {y}_0}
\\
& & 
V_{\tilde{y}_{1}} (t_{k_{1}}) \cdots 
 V_{\tilde{y}_{2n-|l|}} (t_{k_{2n-|l|}})
\prod_{m=1}^n  
  f ( x_{\sigma_m} - x_{\firstpairing_m}, t_{\sigma_m} - t_{\firstpairing_m}, l_{\firstpairing_m})\;.
\eea
Note that a similar formula holds at finite volume for 
$\Vv_{\la,n}^\La ( \underline{\pi} , \tv, \xv, \lv)$.

We must bound the norm of the right-hand side of (\ref{eq-Vv_n}).
Let us denote by $G ( t; x,y ) = \bra{x} \e^{-\I t \HP} \ket{y}$ the 
time-dependent Green function associated to the free motion of the particle
at infinite volume.
Using  (\ref{eq-V_x(t)}) and setting
$t_{j_0} = t_{k_0}=0$, one gets
\begin{equation} \label{eq-intermediatestep}
\bra{x} V_{\tilde{x}_{|l|}} (t_{j_{|l|}}) \cdots 
V_{\tilde{x}_{1}} (t_{j_{1}}) \ket{ {x}_0}
 = 
G (-t_{j_{|l|} }; x,\tilde{x}_{|l|} )   
\mathop{\prod}\limits^{\leftarrow}_{p=1, \ldots, |l|}  W  
 G  ( t_{j_{p}} -  t_{j_{p-1}}; \tilde{x}_{p}, \tilde{x}_{p-1}  ) 
\end{equation}
and
\begin{equation}
\bra{ {y}_0} 
V_{\tilde{y}_{1}} (t_{k_{1}}) \cdots V_{\tilde{y}_{2n-|l|}} (t_{k_{2n-|l|}})
\ket{y}
 = 
\mathop{\prod}\limits^{\rightarrow}_{q=1, \ldots, 2n-|l|} 
G  (t_{k_{q-1}} - t_{k_q};  \tilde{y}_{q-1},\tilde{y}_q ) W 
G ( t_{k_{2n-|l|}}; \tilde{y}_{2n-|l|}, y )  \;.
\end{equation}
Thanks to the propagation bound (\ref{eq: propagation estimate}), we have
\bea \label{eq-bound-Vv}
& & 
\bigl\| 
 \bigl( \Vv_{\la,n} ( \underline{\pi} , \tv,\xv,\lv ) \bigr)_{x_0,y_0;x, y}  
\bigr\|
\\
& & \nonumber
\hspace*{0.3cm} 
\leq    (\| W\| \lambda)^{2n} 
\e^{4  \lambda^2 t} \e^{-|x - \tilde{x}_{|l|} | - |y - \tilde{y}_{2n-|l|} |} 
\prod_{p=1}^{|l|} \e^{- | \tilde{x}_p- \tilde{x}_{p-1} |}  
\prod_{q=1}^{2n - |l|} \e^{- | \tilde{y}_q- \tilde{y}_{q-1} |}  
\prod_{m=1}^n  h_n (t_{\sigma_m}-t_{\firstpairing_m} )\, 
\e^{\frac{| x_{\sigma_m}-x_{\firstpairing_m} |}{2n}} \;.
\eea
Next, observe that 
\be \label{eq-lower_bound_sum_coordinates}
\sum_{p=1}^{|l|} | \tilde{x}_p - \tilde{x}_{p-1} | + 
\sum_{q=1}^{2n-|l|} | \tilde{y}_q - \tilde{y}_{q-1} |
\geq \frac{1}{n} \sum_{m=1}^n |  x_{\sigma_m}-x_{\firstpairing_m}  | - | \tilde{x}_0- \tilde{y}_0 |
\;.
\ee
Actually, (\ref{eq-lower_bound_sum_coordinates}) 
is a consequence of the inequality
\be
\nonumber
\sum_{m=0}^{2n+1} | z_{m+1} - z_{m} |
\;\geq\; 
\max \bigl\{ \bigl| z_\sigma - z_\firstpairing \bigr| \; ;\; 
\firstpairing , \sigma  = 0,\cdots,2n+1 
\bigr\}  
\ee
applied to $(z_0,\cdots , z_{2n+1}) = 
( \tilde{x}_{|l|},\cdots, \tilde{x}_0, \tilde{y}_0,\cdots , \tilde{y}_{2n-|l|} )
\in \integer^{(2n+2)d}$.
Replacing (\ref{eq-lower_bound_sum_coordinates}) into (\ref{eq-bound-Vv}), 
one gets the result  with
\be
R_{x_0,y_0;x,y} (\xv,\lv)=  
\Bigl(2 \sum_{z\in\Zd} \e^{-\frac{1}{2} |z|} \Bigr)^{-2n} 
 \e^{-|x - \tilde{x}_{|l|} | - |y - \tilde{y}_{2n-|l|} |}  \e^{\frac{1}{2} | x_0 - y_0 |}  
\prod_{p=1}^{|l|} \e^{- \frac{1}{2} | \tilde{x}_p- \tilde{x}_{p-1} |}  
\prod_{q=1}^{2n - |l|} \e^{-\frac{1}{2}  | \tilde{y}_q- \tilde{y}_{q-1} |}\;.
\ee 
The proof for the finite lattice is the same, since we only used the propagation 
estimate. 
\finpro

%%%%%%%%%%%%%%%%%%%%%%%%%%%%%%%%%%%%%%%%%%%%%%%%%%%%%%%%%%%%%%%%%%%%%%
\subsection{Uniform convergence of the Dyson series }
\label{sec-thermo_limit} 
%%%%%%%%%%%%%%%%%%%%%%%%%%%%%%%%%%%%%%%%%%%%%%%%%%%%%%%%%%%%%%%%%%%%%%5

%%%%%%%%%%%%%%%%%%%%%%%%%%%%%%%%%%%%%%%%%%%%%%%%%%%%%%%%%%%%%%%%%%%%%%%5
\begin{prop} \label{lem: sum of dyson}
Let $x_0,y_0 \in \Zd$. For fixed $\lambda$ and  $t$, one has 
\be \label{eq-lem sum finite}
\sum_n   \sum_{\xv,\lv}
\sum_{\underline{\pi}}
 \sum_{x,y} \,\mathop{\int}\limits_{Z_{2n}(t)} \D \tv \, \sup_{\Lambda \subset \Zd}  \bigl\| 
\bigl( \Vv_{\lambda,n}^\La (\underline{\pi},\tv,\xv,\lv) \bigr)_{x_0,y_0;x,y}  
\bigr\|   < \infty
\ee
where the supremum is taken over $\La = ]-L,L]^d \cap \Zd$ for all finite $L >0$.
Similarly, assume that the correlation function $f(x,t)$ satisfies \eqref{eq-integrability_assumption} and fix
$\tau \geq 0$, then, 
\be \label{eq-lem sum}
\sum_n 
  \sum_{\xv,\lv}  \sum_{x,y}
\, \sup_{\lambda >0}  \Bigl\{ \sum_{\underline{\pi}}
\mathop{\int}\limits_{Z_{2n}( \lambda^{-2} \tau)} \D \tv \,   \bigl\| 
\bigl( \Vv_{\lambda,n} (\underline{\pi},\tv,\xv,\lv) \bigr)_{x_0,y_0;x,y}  
\bigr\| \Bigr\}  < \infty\;.
\ee
\end{prop}
%%%%%%%%%%%%%%%%%%%%%%%%%%%%%%%%%%%%%%%%%%%%%%%%%%%%%%%%%%%%%%%%%%%%%%%%%%%

\proof
We first show the second claim. The proof of the first one follows similar lines.
We bound (\ref{eq-lem sum}) with the help of Lemma~\ref{lemma1} by
\be \label{eq-lem sum continued}
C_{x_0,y_0} \e^{4 \tau} \sum_n   \sup_{\lambda >0} 
\Bigl\{ (C \lambda)^{2n} \, 
\sum_{\underline{\pi}}
\mathop{\int}\limits_{Z_{2n}( \lambda^{-2} \tau)} \D \tv
\,  
\prod_{m=1}^n h_{n} (t_{\sigma_m}-t_{\firstpairing_m} )
\Bigr\}
 \ee
where $C_{x_0,y_0} < \infty$ denotes the supremum in (\ref{eq-eq-C_x_0_y_0}). 
The sum over all pairings $\underline{\pi}$ and the time integrals are conveniently rewritten with the help of  

%%%%%%%%%%%%%%%%%%%%%%%%%%%%%%%%%%%%%%%%%%%%%%%%%%%%%%%%%%%%%%%%%%%%%%%%%%%%%%
\begin{lemma} \label{lemma2}
For  any locally integrable function
$g : \real^{2n} \rightarrow \real$  and  $0\leq t_0 < t$,
\bea \nonumber
& & 
\sum_{\underline{\pi}}
\mathop{\int}\limits_{t_0 \leq t_1 \leq \cdots \leq t_{2n} \leq t} \D \tv\,
g \bigl( 
t_{\firstpairing_1}, t_{\sigma_1}; \cdots ; t_{\firstpairing_n}, t_{\sigma_n} 
\bigr)
\\[3mm] \nonumber
& & 
\hspace{2cm} 
=
\mathop{\int}\limits_{t_0 \leq {u}_1 \leq \cdots \leq {u}_n \leq t}\!\!\! 
 \D {u}_1 \cdots \D {u}_n 
\mathop{\int}\limits_{ {u}_m \leq {u}_m' \leq t ; m=1,\cdots,n}
\!\!\! \D {u}_1' \cdots \D {u}_n' \, 
g \bigl(
{u}_1,{u}_1' ;\cdots ; {u}_n,{u}_n'  
\bigr)\;.
\eea
\end{lemma}
We leave  to the reader the proof of this lemma, which is based on  change of variables.

%%%%%%%%%%%%%%%%%%%%%%%%%%%%%%%%%%%%%%%%%%%%%%%%%%%%%%%%%%%%%%%%%%%%%%%%%%

\vspace{1mm}

 Applying Lemma \ref{lemma2} with $g(u_1,u_1';\cdots; u_n,u_n')=\prod_m h_n(u_m'-u_m)$ 
as in \eqref{eq-lem sum continued}, bounding the integrals over the $u_m$ and $u_m'$
by $\norm h_n \norm_1^n =(\int_0^\infty \D t\, h_n (t))^n$ times the volume of the $n$-dimensional simplex 
$Z_n(\lambda^{-2} \tau)$, 
we conclude that  \eqref{eq-lem sum continued} is smaller than
\be \label{eq-series_c_n}
C_{x_0,y_0}  \e^{4\tau} 
\sum_{n\geq 0} \frac{1}{n!} \left( {C}^2  \| h_n\|_1 \, \tau \right)^n 
\;.
\ee
By Stirling formula $n ! \sim (2 \pi n)^{1/2} (n/e)^n$ as $n \rightarrow \infty$, one finds that the convergence of 
the series in (\ref{eq-series_c_n}) is ensured by
assumption (\ref{eq-integrability_assumption}), that is, by the condition
$\| h_n\|_1/n \rightarrow 0$ as 
$n \rightarrow \infty$. Thus the second claim of  Proposition~\ref{lem: sum of dyson} is proven.

To show  the first claim, 
we replace  $h_n$ by $\max_{x \in \La} \str f^\La (x,\cdot ) \str$  
(see Lemma \ref{lemma1}) in \eqref{eq-lem sum continued}. Here we do not need to assume that
this function has a finite $L^1$-norm, we bound it  by $f^\La (0,0)$, 
see~(\ref{eq-correl_function}). The quantity in 
 (\ref{eq-lem sum finite}) is thus smaller than 
\begin{equation}
C_{x_0,y_0} e^{4 \lambda^2 t} \sum_{n\geq 0} \frac{1}{n!} \left( \frac{C^2 \lambda^2 t^2 \sup_\La f^\La (0,0)}{2} \right)^n
< \infty \;.
\end{equation}
Note that $\sup_{\La} f^\La (0,0)$ is
finite since $f^\La (0,0)$ converges as $\La \uparrow \Zd$. 
This concludes the proof of Proposition \ref{lem: sum of dyson}.
\finpro

%%%%%%%%%%%%%%%%%%%%%%%%%%%%%%%%%%%%%%%%%%%%%%%%%%%%%%%%%%%%%%%%%%%%%%%
\subsection{Proof of Proposition~\ref{lem: thermo limit}}
\label{sec_proof_thermodyn_limit}
%%%%%%%%%%%%%%%%%%%%%%%%%%%%%%%%%%%%%%%%%%%%%%%%%%%%%%%%%%%%%%%%%%%%%%%

One has
\be \label{eq-pt_limit_single_term_bis}
\bigl\| \bigl(
   (\Vv^{\La}_{\lambda,n} -   \Vv_{\lambda,n})  ( \underline{\pi} , \tv, \xv, \lv) 
 \bigr)_{x_0,y_0;x,y} \bigr\|
\quad \longrightarrow  \quad 0  \qquad \text{as} \quad  \La \nearrow \Zd\;.
 \ee
Actually, let us set $\Aa^\La= (\Vv^{\La}_{\lambda,n} -   \Vv_{\lambda,n})  ( \underline{\pi} , \tv, \xv, \lv)$.
Since $(\Aa^\La)_{x_0,y_0;x,y}$ is a finite matrix, it is sufficient to 
prove that $\| (\Aa^\La)_{x_0,y_0;x,y} (M) \| \leq \| (\Aa^\La)_{x_0,y_0;x,y} (M)\|_1 \rightarrow 0$ for any  
$M \in \caB ( \complex^N)$, so that the convergence 
(\ref{eq-pt_limit_single_term_bis}) follows directly from the last claim in (\ref{eq: three claims}).
By Proposition \ref{lem: sum of dyson}, (\ref{eq-pt_limit_single_term_bis}), and 
dominated convergence,  for any fixed $\la$ and $t$ we have
\be  \label{eq-convergence_propa_int_pict}
\mathop{\ptlim}\limits_{\La \nearrow \Zd}  \left\{\propaDysonLa{t} -   \propaDyson{t} \right\}  =0 \;.
\ee
Since $\propaLa{t}$ preserves the trace and is completely positive, the same holds true for $\propaDysonLa{t}$ 
in (\ref{eq-propa=Dyson_series}). This implies that 
$\| \propaDysonLa{t} \| = 1$ for any $\La$ (see the discussion after (\ref{eq-pt_convergence_scaling_limit})). 
Applying Lemma~\ref{def-lem}, we deduce from  (\ref{eq-convergence_propa_int_pict}) that
\be  \label{eq-convergence_propa_int_pict_strong}
\propaDysonLa{t} (T) \rightarrow    \propaDyson{t} (T)\qquad 
\ee
for any $T \in \traceclass$ with finite support on the lattice $\Zd$. 
Hence $\propaDyson{t}$ can be extended to a bounded, trace-preserving and completely positive 
operator on $\traceclass$.
From this, we straightforwardly deduce that, for a convergent sequence $\rho^{\La}_P \to \rho_P$ in $\traceclass$, 
\be  \label{eq: convergence also for infinite}
\lim_{\La \uparrow \Zd} \propaDysonLa{t} ( \rho^{\La}_P)  =  \propaDyson{t} ( \rho_P )\;.
\ee
Since $\e^{-\I  t [H^{\La}_P, \cdot]} \to \e^{-\I  t [H_P, \cdot]}  $ strongly, 
\eqref{eq: convergence also for infinite} also holds if we replace  
$\propaDysonLa{t}$ by  $\propaLa{t} = \e^{-\I  t [H^{\La}_P, \cdot]}\propaDysonLa{t}$ and 
$\propaDyson{t}$ by $\propa{t}$ which is given by (\ref{eq-link_propagator_and_Dyson_series}).
Therefore, the limit in Proposition~\ref{lem: thermo limit} exists and 
$\propa{t}$  is trace-preserving
and completely positive on $\traceclass$.
\finpro

%%%%%%%%%%%%%%%%%%%%%%%%%%%%%%%%%%%%%%%%%%%%%%%%%%%%%%%%%%%%%%%%%%%%%
\subsection{Crossing diagrams vanish in the van Hove limit}
\label{sec-vanishing_crossed_diagrams}
%%%%%%%%%%%%%%%%%%%%%%%%%%%%%%%%%%%%%%%%%%%%%%%%%%%%%%%%%%%%%%%%%%%%

In this subsection and the following ones, the correlation function $f(x,t)$
does not need to satisfy the assumption (\ref{eq-integrability_assumption}) and we only require that
\be \label{eq: non uniform integrability}
 \norm f(x,\cdot )\norm_1 = \int_0^{\infty} \D t  \str f (x,t) \str     < \infty \qquad 
\textrm{for any}\,\, x \in \bbZ^d\,.
 \ee 
As announced in Step II of the plan of the proof, we show:
 
\vspace{2mm}

%%%%%%%%%%%%%%%%%%%%%%%%%%%%%%%%%%%%%%%%%%%%%%%%%%%%%%%%%%%%%%%%%%%%%%%
\begin{prop} \label{prop-vanishing_crossed_diagrams}
Assume the integrability condition \eqref{eq: non uniform integrability}. 
If $\underline{\pi}$ is a crossing diagram, i.e., $\underline{\pi} \neq \underline{\pi}_{\rm ladder}$, then
for any fixed $n$, $\tau$, $\xv$, $\lv$, and $x_0,y_0,x,y$,
\be  \label{eq: to bound crossing}
\lim_{\la \to 0} \mathop{\int}\limits_{Z_{2n}( \lambda^{-2} \tau)}  \D \tv \,  
\| \Vv_{\lambda,n} ( \underline{\pi} , \tv,\xv,\lv )_{x_0,y_0;x,y} \| =0\;.
\ee
\end{prop}
%%%%%%%%%%%%%%%%%%%%%%%%%%%%%%%%%%%%%%%%%%%%%%%%%%%%%%%%%%%%%

\vspace{3mm}

\proof It is a simple adaptation of the arguments used in~\cite{Wojciech08}, Section 6.3. 
Let
$\underline{\pi} = \{ (\firstpairing_1,\sigma_1),\cdots,(\firstpairing_n,\sigma_n)\}$ be
a crossing diagram. This means that one can find  two pairs
$(\firstpairing_\mu,\sigma_\mu)\in \underline{\pi}$ and $(\firstpairing_\nu,\sigma_\nu)\in \underline{\pi}$
such that $\firstpairing_\mu < \firstpairing_\nu$ (\ie, $\mu < \nu$) and $\firstpairing_\nu < \sigma_\mu$. 
According to (\ref{eq-bound-Vv}), we need to show that
\be
 J_\lambda ( {\underline{\pi}}, \tau) 
 =
\lambda^{2n} \mathop{\int}\limits_{Z_{2n}( \lambda^{-2} \tau)} 
\D \tv \prod_{m=1}^n  \str f_{m} (  t_{\sigma_m} - t_{\firstpairing_m}  )  \str
\quad \longrightarrow \quad 0\;
\ee
as $\lambda \rightarrow 0$, where we
abbreviated  $f ( x_{\sigma_m} - x_{\firstpairing_m}, t_{\sigma_m} - t_{\firstpairing_m})$ 
by  $f_m(t_{\sigma_m} - t_{\firstpairing_m})$ (recall that here $\xv$ is fixed). One has 
\bea
\nonumber
& &   J_\lambda ( {\underline{\pi}}, \tau) 
 \leq  
\prod_{m\not= \mu,\nu}^n \int_{0 \leq t_{\firstpairing_m} \leq t_{\sigma_m} \leq \lambda^{-2} \tau}
\D t_{\firstpairing_m}\D t_{\sigma_m} \, \lambda^2 \str f_{m} (  t_{\sigma_m}- t_{\firstpairing_m} ) \str
\\ \nonumber
& &
\hspace*{1cm} \times 
\int_{0 \leq t_{\firstpairing_\mu} \leq t_{\firstpairing_\nu} \leq t_{\sigma_\mu} \leq \lambda^{-2} \tau, 
t_{\firstpairing_\nu} \leq t_{\sigma_\nu}} 
\D t_{\firstpairing_\mu} \D t_{\sigma_\mu} \D t_{\firstpairing_\nu} \D t_{\sigma_\nu} \,
 \lambda^4 \str f_{\mu} ( t_{\sigma_\mu} - t_{\firstpairing_\mu} ) \str \str  f_{\nu} ( t_{\sigma_\nu}- t_{\firstpairing_\nu} ) 
\str\;.
\eea
The first  (product of) integrals on the right-hand side is bounded by  
$( \sup_m \| f_{m} \|_1 \tau)^{n-2}$.  
To deal with the last integral, we first bound the integral over $t_{\sigma_\nu}$ by 
 $\lambda^4 \norm f_{\nu} \norm_1 f_\mu ( t_{\sigma_\mu} - t_{\firstpairing_\mu} )$ and then  
substitute  $v=\lambda^2 t_{\sigma_\mu}$, 
$w=\lambda^2 (t_{\sigma_\mu} - t_{\firstpairing_\nu} )$, and $t'=t_{\sigma_\mu}-t_{\firstpairing_\mu}$.
This gives the bound
\be \label{eq-maj_I}
\norm f_{\nu} \norm_1 
\int_0^\tau \D v \int_0^v \D w \int_{\lambda^{-2} w}^{\lambda^{-2} v} \D t' \,  \str f_{\mu} (t')\str \;.
\ee
For any $(v,w) \in [0,\tau]^2$ such that $0< w < v$, 
$\int_{\lambda^{-2} w}^{\lambda^{-2} v} \D t' \, \str f_{\mu} (t')\str$ converges to zero as
$\lambda \rightarrow 0$ (because $\| f_\mu \|_1 < \infty$).
This integral is also bounded by $\| f_\mu \|_1$, therefore 
(\ref{eq-maj_I}) converges to zero by dominated convergence.
\finpro

%%%%%%%%%%%%%%%%%%%%%%%%%%%%%%%%%%%%%%%%%%%%%%%%%%%%%%%%%%%%%%%%%%%%%
\subsection{Contribution of the ladder diagrams}
\label{sec-ladder_diagrams}
%%%%%%%%%%%%%%%%%%%%%%%%%%%%%%%%%%%%%%%%%%%%%%%%%%%%%%%%%%%%%%%%%%%%

In this subsection, we determine the contribution of the ladder diagrams and
accomplish Step III of the proof.  
We first introduce the following operators on $\traceclass$:
\be
\caU_\la(t) =  \e^{-i t [H_P, \cdot]} =   \e^{-i t [\la^2H_{\mathrm{hop}}+S, \cdot]}   
\qquad , \qquad \caU_0(t) =    \e^{-i t [S, \cdot]}  \;.
\ee
Define the family of operators 
\be
\mathcal{A}(x',l'; x,l) = - 
\int_0^\infty \D t\,      
f(x'-x,t, l)      \caU_0(-t)  \Ii ({x'},0,l')  \caU_0(t)  \Ii (x,0,l)\;.
\ee
The integral is convergent by the integrability condition (\ref{eq: non uniform integrability}). 
Let 
\be  \label{def: caw}
 \Ww_{\lambda,n} (\underline\tau,\xv,\lv) 
 =   \mathop{\prod}\limits_{j=1,\ldots,n}^{\leftarrow}     \mathcal{U}_\la (-\lambda^{-2} \tau_{j} ) 
 \mathcal{A}(x_{2j},l_{2j};x_{2j-1},l_{2j-1})  \mathcal{U}_\la (\lambda^{-2} \tau_{j} )
\ee
for any $\underline\tau=(\tau_1, \ldots, \tau_n) \in \real_+^n$.  
 We have

%%%%%%%%%%%%%%%%%%%%%%%%%%%%%%%%%%%%%%%%%%%%%%%%%%%%%%%%%%%%%%%%%
\begin{prop} \label{prop-ladder_diagrams} 
Assume the integrability  condition \eqref{eq: non uniform integrability}.
For any fixed $\xv$, $\tv$ and $\tau$,
\be \label{eq-prop-2} 
\lim_{\la \to 0}  \;
\Bigl\| \mathop{\int}\limits_{Z_{2n}( \lambda^{-2} \tau)} \D \tv   \, 
\Vv_{\lambda,n} ( \underline{\pi}_{\rm ladder} , \tv, \xv, \lv) 
- \mathop{\int}\limits_{Z_n(\tau)} \D \underline\tau  \,   \Ww_{\lambda,n}(\underline{\tau},\xv,\lv)  
\Bigr\| =0\;.
\ee
\end{prop}
%%%%%%%%%%%%%%%%%%%%%%%%%%%%%%%%%%%%%%%%%%%%%%%%%%%%%%%%%%%%%%%%%%%

\proof 
Let us set, for any $\delta>0$, 
\be
\Aa_{\lambda,\delta} (x',l';x,l) = 
- \int_0^{\lambda^{-2} \delta}  \D t  \,     f(x'-x,t, l)   
\mathcal{U}_\lambda (-t)\Ii ({x'},0,l')  \mathcal{U}_\lambda (t)  \Ii ({x},0,l)\;.
\ee
Then
\be \label{eq: convergence caH}
\lim_{\la \to 0} \Aa_{\lambda,\delta } ( x',l';x,l)
 =   \Aa (x',l';x,l)
\; \text{ in norm and } \; 
\| \Aa_{\lambda,\delta } ( x',l';x,l) \| \leq \| f(x'-x,\cdot)\|_1 \| W\|^2
\ee
for any $x,x' \in \Zd$, and $l,l'\in \{ L,R\}$. This follows from the dominated convergence theorem, using 
(i)~the integrability of $\str f(x'-x,\cdot) \str$, (ii)~the norm convergence 
 $\lim_{\la\to 0}  \mathcal{U}_\lambda (t) = \mathcal{U}_0 (t)$ 
(which follows directly from the boundedness of $\Hkin$), and (iii)~the bounds 
 $\| \Ii (x,0,l) \| \leq \| W\|$ and  $\norm \mathcal{U}_\lambda(t) \norm = 1$.
Now,  using 
$\Ii (x,t,l) = \mathcal{U}_\la (- t) \mathcal{I} (x,0,l)  \mathcal{U}_\la ( t)$
and setting $s_j = t_{2j}-t_{2j-1}$, we rewrite the (infinite volume version of) (\ref{eq-def-Vv}) as 
\be  
\begin{split}
 \Vv_{\lambda,n} (  \underline{\pi}_{\rm ladder} , \tv, \xv, \lv)  
 = &  (-\lambda^2)^n  \mathop{\prod}\limits_{j=1,\ldots,n}^{\leftarrow}   f(x_{2j}-x_{2j-1}, s_j,l_{2j-1})  
\mathcal{U}_\la (-t_{2j-1})     \\  
& 
\mathcal{U}_\la (-s_j)\Ii (x_{2j}, 0, l_{2j})  \mathcal{U}_\la (s_j)  \Ii (x_{2j-1}, 0, l_{2j-1})   
 \mathcal{U}_\la (t_{2j-1})  \;.  
\end{split}
\ee
We now perform the variable substitutions $\tau_j = \lambda^{2} t_{2j-1}$, $s_j = t_{2j}-t_{2j-1}$ for
$j=1,\cdots,n$, to get
\begin{eqnarray}
& & 
\mathop{\int}\limits_{Z_{2n}(\la^{-2} \tau)} \D \tv \,
\Vv_{\lambda,n} (  \underline{\pi}_{\rm ladder} , \tv, \xv, \lv)  
- \mathop{\int}\limits_{Z_{n}(\tau)} \D \tauv \,
\Ww_{\lambda,n} ( \tauv, \xv, \lv)  
=\mathop{\int}\limits_{Z_{n}(\tau)} \D \tauv \,
\Bigl\{
 \mathop{\prod}\limits_{j=1,\ldots,n}^{\leftarrow} \mathcal{U}_\la (-\frac{\tau_j}{\lambda^2})  
\\
\nonumber
& &
\hspace*{0.6cm} 
  \Aa_{\lambda,\tau_{j+1} - \tau_j} ( x_{2j}, l_{2j}; x_{2j-1},l_{2j-1} ) 
   \mathcal{U}_\la (\frac{\tau_j}{\lambda^2}) 
- \mathop{\prod}\limits_{j=1,\ldots,n}^{\leftarrow} 
\mathcal{U}_\la (-\frac{\tau_j}{\lambda^2}) 
  \Aa ( x_{2j}, l_{2j}; x_{2j-1},l_{2j-1} ) 
   \mathcal{U}_\la (\frac{\tau_j}{\lambda^2}) 
\Bigr\}
\;.
\end{eqnarray}
For any $0 \leq \tau_1 < \tau_2 < \cdots < \tau_n$, the integrand inside the curly brackets
converges in norm to zero because of (\ref{eq: convergence caH}).
This integrand is bounded by $2 ( \max_{j=1,\cdots,n} \|  f(x_{2j}-x_{2j-1},\cdot)\|_1 \| W\|^2 )^n$. 
Hence an application of the 
dominated convergence theorem yields the result.  
\finpro

%%%%%%%%%%%%%%%%%%%%%%%%%%%%%%%%%%%%%%%%%%%%%%%%%%%%%%%%%%%%%%%%%%%%%%%%%%%
\subsection{Spectral averaging} \label{sec: spectral averaging}
%%%%%%%%%%%%%%%%%%%%%%%%%%%%%%%%%%%%%%%%%%%%%%%%%%%%%%%%%%%%%%%%%%%%%%%%%%
%
To end the proof of Theorem~\ref{thm: main}, we use some standard techniques 
of ``dynamical spectral averaging'', originally used in Ref.~\cite{Davies76} in the same context. 

\vspace{1mm}

%%%%%%%%%%%%%%%%%%%%%%%%%%%%%%%%%%%%%%%%%%%%%%%%%%%%%%%%%%%%%%%%%%%%%%%%%%%%%%%%%
\begin{lemma} \label{lem: spectral averaging}
Let $A$ and $B$ be bounded operators on  a Banach space $\mathcal Y$ such that  $(\e^{t B})_{t \in \real}$ is a
 one-parameter group of isometries on $\mathcal Y$ and the norm limit
\be  \label{eq: general sharp}
A^{{\natural}} = \lim_{t \to \infty} t^{-1}   \int_0^t d u  \,  \e^{-u B}  A \e^{u B}
\ee
exists.  Let $D(\cdot),E(\cdot)$ be in $\caC^1(\mathbb{R}, \mathcal{B}(\mathcal Y))$ 
(continuously differentiable $\mathcal{B}(\mathcal Y)$-valued functions). Then, for any $\tau >0$, 
\begin{enumerate}
\item[(1)] $\displaystyle\quad  
\lim_{\varepsilon \to 0} \int_0^\tau \D \tau_1 \,D(\tau_1) \e^{-(\tau_1/\varepsilon) B} A \e^{(\tau_1/\varepsilon) B} E(\tau_1) 
 = \int_0^\tau \D \tau_1 \, D(\tau_1) A^{{\natural}} E(\tau_1)$
\item[(2)] $\displaystyle  \quad 
\lim_{\varepsilon \to 0}  \e^{- (\tau/\varepsilon) B} \e^{ (\tau/\varepsilon)( B+ \varepsilon A)}
 =  \lim_{\varepsilon \to 0}   \e^{ (\tau/\varepsilon)( B+ \varepsilon A)} \e^{- (\tau/\epsilon) B} 
 = \quad \e^{ \tau   A^{{\natural}}} $.
\end{enumerate}
\end{lemma}
%%%%%%%%%%%%%%%%%%%%%%%%%%%%%%%%%%%%%%%%%%%%%%%%%%%%%%%%%%%%%%%%%%%%%%

\vspace{1mm}

\proof
To show the claim {\it (1)}, we put $E(\tau)=1$ (the general result follows by an obvious extension of the proof). 
Let us write $D'(\tau) = \frac{d }{d \tau}D(\tau)$,  then
\begin{eqnarray}
& & \int_0^\tau \D \tau_1      D(\tau_1)  \e^{-(\tau_1/\varepsilon) B}  A \e^{(\tau_1/\varepsilon) B} 
 =     
\int_0^\tau \D \tau_1 \left( D(0)+ \int_0^{\tau_1} \D \tau_2   \, D'(\tau_2) \right) \e^{-(\tau_1/\varepsilon) B}  
A \e^{(\tau_1/\varepsilon) B}  \nonumber 
\\ 
 & & \hspace*{2cm}  =     
\int_0^\tau \D \tau_1  \,  D(0) \e^{-(\tau_1/\varepsilon) B}  A \e^{(\tau_1/\varepsilon) B}  +    \int_0^\tau \D \tau_2  \,    
D'(\tau_2) \int_{\tau_2}^\tau \D \tau_1 \, \e^{-(\tau_1/\varepsilon) B}  A \e^{ (\tau_1/\varepsilon) B}  \nonumber   
\\
&  & \hspace*{2cm} \mathop{\rightarrow}\limits_{\varepsilon \to 0}     
\left(\tau D(0) + \int_0^\tau \D \tau_2  \, (\tau-\tau_2) D'(\tau_2) \right)  A^{{\natural}}   
 = \int_0^\tau \D \tau_1 \, D(\tau_1) A^{{\natural}} \;.
\end{eqnarray}
 To get the 
last line, we used \eqref{eq: general sharp} to estimate the integrals over $\tau_1$ for all $\tau_2<\tau$, 
together with the
dominated convergence theorem (since  $ D'(\cdot)$ is norm continuous and $\e^{t B}$ is an isometry). 

The claim {\it (2)} can be proven from {\it (1)} by expanding the two first members as Dyson series in $A$ 
(alternatively, see~\cite{Davies_book_semigroup}, Theorem 5.11,  
and the review by Derezi\'{n}ski and Fruboes in~\cite{attalreview} for the same result under weaker conditions). 
\finpro

\vspace{4mm}

Let us now apply Lemma~\ref{lem: spectral averaging} to the case at hand.  We choose $B= \I [S,\cdot]$, 
$\varepsilon=\la^2$ and $\mathcal Y= \traceclass$.  Then 
$( \e^{t B})_{t \in \real}$
is a group of isometries on $\traceclass$.
Since $S$ has a finite number of eigenvalues, the existence of the norm limit \eqref{eq: general sharp} is 
automatic. Moreover,  for any $\caA \in \caB(\traceclass)$ one has
\be
\caA^\natural 
 =  
\sum_{\omega \in \spec([S,\cdot])}
 {\cal P}_\omega  \caA
{\cal P}_\omega
\ee
where  ${\cal P}_\omega$ are the spectral projectors of 
$[ S,\cdot]$, i.e.,
\be  \label{eq-projector_spectral_av}
\Pp_\omega ( T )= \sum_{s,s'=1,\cdots, N} \delta_{E_s-E_{s'}, \omega} \ketbra{s}{s'} \bra{s} T \ket{s'}
\quad , \quad T \in \traceclass  \;.
\ee
We first prove 
%%%%%%%%%%%%%%%%%%%%%%%%%%%%%%%%%%%%%%%%%%%%%%%%%%%%%%%%%%%%%%%%%%%%%%%
\begin{prop} \label{prop_spectral_averaging}
Let us define
\be
 \Kk_{n}(\underline{\tau},\xv,\lv) = 
\prod_{j=1, \ldots, n}^\leftarrow    
\caU_\natural (-\tau_j)   \bigl[ \Aa ( x_{2j},l_{2j}; x_{2j-1},l_{2j-1}) \bigr]^{\natural}  \,   \caU_\natural (\tau_j) 
\qquad \text{ with } \quad \caU_\natural (\tau) =    \e^{ -i \tau  [H^\natural_{\mathrm{hop}}, \cdot]}
\ee
where $H^\natural_{\mathrm{hop}}$ is given by 
(\ref{eq-spectrally_av_hopping_Hamil}).
Then for the norm topology on $\Bb (\traceclass)$,
\be
\lim_{\la \to 0}   \mathop{ \int}\limits_{Z_n(\tau)} \D \underline{\tau}  \,
 \Ww_{\lambda,n}(\underline{\tau},\xv,\lv)  
=     \mathop{ \int}\limits_{Z_n (\tau)}  \D \underline{\tau} \Kk_{n}(\underline{\tau},\xv,\lv)\;.
\ee
\end{prop}
%%%%%%%%%%%%%%%%%%%%%%%%%%%%%%%%%%%%%%%%%%%%%%%%%%%%%%%%%%%%%%%%%%%%%%%%%%%%%%%%%%%%%%

\vspace{1mm}

\proof
An explicit calculation yields 
\be
[ \Hkin , \cdot ]^\natural = [ \Hkin^\natural , \cdot ]\;.
\ee
Choosing $A= \I [H_{\mathrm{hop}}. \cdot]$,
 the claim {\it (2)} of Lemma \ref{lem: spectral averaging} yields
\be \label{eq: spectral av to cau}
\lim_{\la \to 0}  \|    \caU_0(-\tau/\la^2)\, \caU_\lambda (\tau/\la^2) -  \caU_\natural (\tau) \| =
\lim_{\la \to 0}  \|    \caU_\lambda (\tau/\la^2)\,  \caU_0(-\tau/\la^2) -  \caU_\natural (\tau) \| =
0 \;.
\ee
We use the  abbreviation $\caA_j = \Aa (x_{2j},l_{2j}; x_{2j-1},l_{2j-1})$.
 Since ${\cal U}_0 (t)$ is an isometry, this gives
\be \label{eq: spectral av to cau_bis}
\lim_{\la \to 0} 
\| 
\caU_\lambda ( -\tau/\la^2)  \Aa_j \caU_\lambda ( \tau/\la^2) 
 -
  \caU_\natural (-\tau) \caU_0 (-\tau/\la^2 ) \Aa_j \caU_0 (\tau/\la^2 ) \caU_\natural (\tau) \| = 0\;.
\ee
To prove the Proposition, we consider first the cases $n=1$ and $n=2$ and then conclude by induction.

{For $n=1$,} one has   
\begin{eqnarray}
& &  \lim_{\la \to 0}   \int_{Z_1(\tau)} \D \underline{\tau} \, \Ww_{1,\la} (\underline{\tau},\xv,\lv)
 =  
\lim_{\la \to 0}   \int_0^\tau \D \tau_1 \,  \caU_\la (-\tau_1/\la^2)    \caA_1    \caU_\la
(\tau_1/\la^2)  \\
\nonumber
& & \hspace*{1cm} 
= \lim_{\la \to 0}   \int_0^\tau \D \tau_1  \,  \caU_\natural (-\tau_1)  \caU_0(-\tau_1/\la^2)
 \caA_1 \caU_0(\tau_1/\la^2)    \caU_\natural (\tau_1)     
=    \int_0^\tau \D \tau_1   \,  \caU_\natural (-\tau_1)  \caA^{\natural}_1  \caU_\natural (\tau_1)     
\end{eqnarray}
where we used \eqref{eq: spectral av to cau_bis} in the second equality, relying also on dominated convergence and
the uniform boundedness of the integrand,  and claim {\it (1)} of Lemma \ref{lem: spectral averaging} 
in the third equality (note that $\caU_\natural (\cdot )$ is $C^1$).

{For $n=2$, one has} 
\begin{eqnarray}
\nonumber
& & \lim_{\la \to 0}   \int_{Z_2(\tau)} \D \underline{\tau} \, \Ww_{2,\la} (\underline{\tau},\xv,\lv)
 =  
\lim_{\la \to 0}   \int_0^\tau \D \tau_2 \int_0^{\tau_2 } \D  \tau_1 \,   \caU_\la (-\tau_2/\la^2)    
\caA_2    \caU_\la (\tau_2/\la^2) \caU_\la (-\tau_1/\la^2)    \caA_1    \caU_\la (\tau_1/\la^2) 
 \\
\nonumber
& & \hspace*{2cm} =    
\lim_{\la \to 0}   \int_0^\tau \D \tau_2 \,   \caU_\la (-\tau_2/\la^2)    \caA_2    \caU_\la (\tau_2/\la^2)   
\int_0^{\tau_2} \D \tau_1\,  \caU_\natural (-\tau_1)  \caA^{\natural}_1  \caU_\natural (\tau_1)       
\\
\nonumber
& & \hspace*{2cm} =    \lim_{\la \to 0}   \int_0^\tau \D \tau_2  \, \caU_\natural (-\tau_2)  \caU_0 ( - \tau_2/\la^2) 
  \caA_2  \caU_0 (\tau_2/\la^2 )  \caU_\natural (\tau_2)   
   \int_0^{\tau_2} \D \tau_1 \,   \caU_\natural (-\tau_1)  \caA^{\natural}_1  \caU_\natural (\tau_1)    
 \\
& & \hspace*{2cm} =    
\int_0^\tau \D \tau_2 \,  \caU_\natural (-\tau_2)    \caA^{\natural}_2   \caU_\natural (\tau_2)   
   \int_0^{\tau_2} \D \tau_1 \,\caU_\natural (-\tau_1)  \caA^{\natural}_1  \caU_\natural (\tau_1)  \;.  
\end{eqnarray}
The second equality is the case $n=1$, the third is \eqref{eq: spectral av to cau_bis},
and the fourth follows from the claim {\it (1)} of Lemma \ref{lem: spectral averaging}.

{The case $n>2$} follows   by a similar induction step. 
\finpro

\vspace{2mm}

\noindent
{\bf End of the proof of Theorem~\ref{thm: main}.} 
Collecting 
Propositions~\ref{prop-vanishing_crossed_diagrams}, \ref{prop-ladder_diagrams}, 
and \ref{prop_spectral_averaging}, we have 
\be
\lim_{\la \to 0} \sum_{\underline{\pi}} \mathop{\int}\limits_{Z_{2n}( \lambda^{-2} \tau)} d \tv   \, 
\bigl( \Vv_{\lambda,n} ( \underline{\pi}, \tv, \xv, \lv) \bigr)_{x_0,y_0;x,y}
=
 \mathop{ \int}\limits_{Z_n(\tau)} \D \underline{\tau}  \,
 \bigl( \Kk_{n}(\underline{\tau},\xv,\lv) \bigr)_{x_0,y_0;x,y}
\ee
for any fixed $n$, $\xv$, $\lv$, and $x,y,x_0,y_0$.  
Thanks to the second statement of Proposition~\ref{lem: sum of dyson} and to the dominated convergence
theorem, we obtain in view of (\ref{eq-Dyson_expansion_therm_limit})
\be \label{eq-almost_finished}
\ptlim_{\la \to 0} 
\propaDyson{\lambda^{-2} \tau} 
= 
\sum_n \sum_{\xv,\lv   } \mathop{ \int}\limits_{Z_n(\tau)} \D \underline{\tau}  \,
\Kk_{n}(\underline{\tau},\xv,\lv) 
= \e^{\I \tau [ \Hkin^\natural, \cdot ]} \e^{-\I \tau [ \Hkin^\natural, \cdot ] + \tau \Aa^\natural }
\ee
where the last equality comes from a Dyson expansion in powers of $\Aa^\natural$ 
(the series is convergent in the $\ptlim$ sense by dominated convergence), and
\be
\Aa^\natural  = \sum_{x,x',l,l'} \bigl[ \Aa ( x,l;x',l') \bigr]^\natural\;.
\ee
One concludes by invoking Lemma~\ref{def-lem} (see also the discussion after this lemma) that 
\be
\propaDyson{\lambda^{-2} \tau} \;\rightarrow \;
 \e^{\I \tau [ \Hkin^\natural, \cdot ]} \e^{-\I \tau [ \Hkin^\natural, \cdot ] + \tau \Aa^\natural }
\qquad \text{ strongly as $\lambda \rightarrow 0$.}
\ee
But $\e^{\I t [ S,\cdot ]}$ and $\e^{- \I t [ \HP ,\cdot] }$ are isometries on $\traceclass$. Therefore, 
in view of (\ref{eq: limit of main result}) and
(\ref{eq-link_propagator_and_Dyson_series}),
\be
\rho_{\rm sl} (\tau) 
= \lim_{\la \to 0}
\e^{\I \la^{-2} \tau [ S,\cdot ]} \e^{- \I \la^{-2} \tau [ \la^2 \Hkin+ S,\cdot] }
\propaDyson{\la^{-2} \tau} (\rho_P)
= \e^{-\I \tau [ \Hkin^\natural, \cdot ] + \tau \Aa^\natural } (\rho_P)
\ee
in the trace-norm topology, where
we have used again Lemma~\ref{lem: spectral averaging} {\it (2)}.

To  establish the agreement with the generator $\caL^\natural$ given by (\ref{eq-Lindblad_generator}), 
we check by inspection that, for any $\rhoP \in \traceclass$,
\begin{eqnarray}
\bigl[ \mathcal{A}(x,l=\links; y,l'=\rechts ) \bigr]^\natural  (\rhoP )
&= &   
 \sum_{\omega \in \spec ([\cdot, S]) }   c(y-x,\omega)    W_{\omega}\otimes \ketbra{x}{x} \,  \rhoP \, 
   W_{\omega}^\ast \otimes \ketbra{y}{y} 
\\[2mm]
\bigl[ \mathcal{A}(x,l=\links; y,l'=L ) \bigr] ^\natural  ( \rhoP ) 
&= & 
\I \delta_{x,y} \Upsilon \rhoP +      \delta_{x,y}   \sum_{\omega \in \spec ([\cdot, S]) }   c(0,\omega) 
    W_{\omega} W_{\omega}^\ast \otimes \ketbra{x}{x} \,  \rhoP \, 
\end{eqnarray}
and 
\be
\bigl[ \mathcal{A}(x,\rechts; y,\links ) \bigr]^\natural( \rhoP )
= \Bigl( \bigl[ \mathcal{A}(x,\links; y,\rechts ) \bigr]^\natural ( \rhoP) \Bigr)^* 
\qquad , \quad  \bigl[ \mathcal{A}(x,\rechts; y,\rechts ) \bigr]^\natural ( \rhoP ) 
 = \Bigl( \bigl[ \mathcal{A}(x,\links; y,\links )\bigr] ^\natural  (\rhoP ) \Bigr)^* \;.
\ee
Consequently, we have
\be
\mathcal{A}^\natural ( \rhoP) =  i [\Upsilon, \rhoP ] +  \mathfrak{A}(\rhoP)- \frac{1}{2} \{ \mathfrak{A}^{\star}(1),\rhoP \} 
\ee
with $\mathfrak{A}$ as defined in \eqref{eq: manifest lindblad form}.  
\finpro

\vspace{5mm}

\noindent {\bf Acknowledgements:}
D.S. is grateful to A. Joye for stimulating discussions and 
acknowledges financial support from
the french Agence Nationale de la Recherche, project ANR-09-BLAN-0098-01.

\appendix
\renewcommand{\theequation}{\Alph{section}\arabic{equation}}
\setcounter{equation}{0}
%%%%%%%%%%%%%%%%%%%%%%%%%%%%%%%%%%%%%%%%%%%%%%%%%%%%%%%%%%%%%%%%%%%%
\section{Appendix: proof of Proposition~\ref{prop-decay_correl_funct} } \label{app-C}
%%%%%%%%%%%%%%%%%%%%%%%%%%%%%%%%%%%%%%%%%%%%%%%%%%%%%%%%%%%%%%%%%%%%%%%%%%

We show in this appendix that the main hypothesis (\ref{eq-integrability_assumption}) 
of Theorem~\ref{thm: main} concerning the decay of the bath correlation functions
$f_i(x,t)$ is satisfied under the following conditions:

\begin{enumerate}

\item[(i)] $d \geq 2$;

\item[(ii)]  the support of $g_{0,i}(q)$ belongs to the open ball
$B_r= \{ q \in \torus^d ; |q| < r\}$ with $0<r\leq \pi$; 
the  form factor $g_{0,i}(q)$ and the momentum occupation numbers $\zeta_i (q)$ 
depend only on $|q|$ on $B_r$;

\item[(iii)] the bosons have a linear dispersion relation: $\nu_i(q) =  \str q \str $ for $q \in B_r$;

\item[(iv)] the non-negative functions
\be \label{eq-def_psi_pm}
\psi_{i,+} (|q| ) = \str g_{0,i}(q) \str^2  \zeta_i (q)
\quad , \quad 
\psi_{i,-} (|q|)= \str g_{0,i} (q)\str^2  (1+ \zeta_i (q))
\ee
belong to $C^2 (]0, \pi])$ and the three functions of $|q|$ below are in $L^1 ([0,\pi])$: 
\be
|q|^{\min\{d-3,\frac{d-1}{2} \} } \psi_{i,\pm} (|q|)
\; , \quad 
 |q|^{d-2} \psi_{i,\pm} ' (|q|) 
\; , \quad  
|q|^{d-1} \psi_{i,\pm} '' (|q|)
\;.
\ee

\end{enumerate}

For simplicity we omit the index $i$ labelling the baths.
Using the notation (\ref{eq-def_psi_pm}), the correlation function (\ref{eq-assumption}) reads 
\be \label{eq-assumptionbis} 
f(x,t) = f_+ (x,t) + f_- (x,t) = \int_{\torus^d} \frac{\D^d q}{(2\pi)^d} 
\left(  \psi_+ (|q|)  \e^{i q \cdot x} \e^{i t \nu (q)}  + \psi_-(|q|)  \e^{-i q \cdot x} 
\e^{-i t \nu (q)  }    \right) \;.
\ee
We first show that under conditions (i-iv),  there exists a 
 constant $C_d>0$ such that for any $n \in \spinteger$,
\be \label{eq-region_without_stat_pts}
 \int_1^\infty \D t \,
 \sup_{x \in \Zd, |x| \geq t/2} \, | f(x,t)| \e^{-\frac{ | x|}{n} }
  \leq C_d
\quad \text{ if $\;d\geq 3$}
\quad , \quad 
 \int_1^\infty \D t \,
 \sup_{x \in \Zd , |x| \geq t/2} \, | f(x,t)| \e^{-\frac{ | x|}{n} }
  \leq C_2 \sqrt{n} 
\quad \text{ if $\;d = 2$.}
\ee
Actually, by (\ref{eq-assumptionbis}) and (ii-iii),  $f(x,t)$ can be rewritten 
as the Fourier transform
\be \label{eq-f(x,t)_as_Fourier_transfo}
f(x,t) = \frac{1}{(2\pi)^{\frac{d}{2}} |x|^{\frac{d-2}{2}} }
\int_{-r}^r \D \omega\, |\omega|^{\frac{d}{2}} \psi_{{\rm{sign}}(\omega)} (|\omega|) 
 J_{\frac{d-2}{2}} ( |\omega x |) \e^{i t \omega}
\ee
where ${\rm{sign}} (\omega)=\pm 1$ for $\pm \omega >0$ and 
$J_{m} (r)=(r/2)^m \Gamma (m+\frac{1}{2})^{-1}( 2/ \sqrt{\pi}) \int_{0}^1 
\D u \, (1-u^2)^{m-\frac{1}{2}} \cos (r u)$ 
denotes the Bessel function of order $m$ (here $\Gamma$ is the Gamma function). A standard bound, 
see e.g.~\cite{Wang}, yields  $C= \sup_{r\geq 0}\{ \sqrt{r} | J_m (r)| \} <\infty$. Hence
\be \label{eq-bound_on_f}
\sup_{|x| \geq t/2}
\bigl| f(x,t) \bigr| 
 \leq 
  \frac{C}{2^{\frac{1}{2}} \pi^{\frac{d}{2}} t^{\frac{d-1}{2} } }
   \int_{-r}^r \D \omega \,|\omega|^{\frac{d-1}{2}} \psi_{{\rm{sign}}(\omega)} (|\omega|)
\;.
\ee 
The last integral is convergent thanks to assumption (iv).
Then (\ref{eq-region_without_stat_pts}) follows from 
\be \label{eq-convergent_integral}
\sup_{n \in \spinteger} 
\int_1^\infty \D t\,  t^{-\frac{d-1}{2}}\, \e^{-\frac{t}{2 n}} < \infty
\quad \text{ if $\;d \geq 3$}
\quad , \quad 
\sup_{n \in \spinteger} 
\frac{1}{\sqrt{n}} \int_1^\infty \D t\,  t^{-\frac{d-1}{2}}\, \e^{-\frac{t}{2 n}} < \infty
\quad \text{ if $\;d = 2$.}
\ee

We now show that there exists a constant $C>0$ such that
\be \label{eq-bound_on_max_of_f}
\max_{x \in \Zd, |x| < t/2} | f(x,t) | \leq \frac{C}{t^2}\;.
\ee
Let us set
\be
v (q,x,t)= \nabla \nu (q) + \frac{x}{t}
\ee
where $\nabla$ is the gradient with respect to $q$.
Note that $v(q,x,t)$ is the gradient of the phase $S (q,x,t)= \nu (q) + q \cdot x/t$
appearing in the oscillatory integral (\ref{eq-assumptionbis}).
Assuming $|x | < t/2$, one has  $| v(q,x,t) | > 1/2$ 
(since $\nabla \nu (q) = q/|q|$ has norm one), thus 
this phase has no stationary points. Noting that
$\e^{i t  S ( q,x,t)} = (i t)^{-1} (v/|v|^2) \cdot \nabla \e^{i t S ( q,x,t)}$ and
integrating twice by part yields 
\be \label{eq-bound_on _I(x,t)}
| f_\pm (x,t) | 
 \leq 
  \frac{1}{t^2} \sum_{k,l=1}^d \int_{\torus^d} \frac{\D^d q}{(2\pi)^d}
   \biggl| 
    \partial_k
     \biggl( \frac{v_k}{|v|^2}
      \partial_l \Bigl( \frac{v_l \,\psi_\pm}{|v|^2} \Bigr)
     \biggr)
   \biggr|
\ee
where $\partial_k$ is the derivative with respect to the $k$th component of $q$.
Note that the boundary terms vanish thanks to condition (ii).
A simple calculation shows that the integrand in the right-hand side 
of (\ref{eq-bound_on _I(x,t)}) is bounded by 
\be
c_1 |q|^{-2} \psi_\pm (|q|)
+ c_2
 |q|^{-1} | \nabla  \psi_\pm (|q|) |
  + c_3 
\max_{k,l=1,\cdots,d} 
\bigl| \partial_k \partial_l  \psi_\pm (|q|) \bigr|
\ee 
for some constants $c_1,c_2$, and $c_3>0$.
The last function is integrable by assumption (iv) and this proves our claim 
(\ref{eq-bound_on_max_of_f}).  

Collecting the above results and recalling that $|f(x,t)| \leq f(0,0)$, we conclude that  
\be
\frac{1}{n} \int_0^\infty \D t \,
 \sup_{x \in \Zd} \, | f(x,t)| \e^{-\frac{ | x|}{n} }
  \leq \frac{1}{n} f(0,0)
+ \frac{1}{n} \int_1^\infty \D t \,\sup_{x,|x| \geq t/2} \, | f(x,t)| \e^{-\frac{ | x|}{n} }
  + \frac{1}{n} \int_1^\infty \D t \,
 \max_{x,|x| <  t/2} \, | f(x,t)| 
\ee
converges to zero as $n \rightarrow \infty$.
This proves
Proposition~\ref{prop-decay_correl_funct}. \finpro

\vspace{2mm}

Let us stress that conditions (i-iv) are not optimal for 
the hypothesis (\ref{eq-integrability_assumption}) to hold.
In particular, the rotation invariance of $g_0$ and $\zeta$ in (ii) and the linear dispersion (iii)
%, which is in fact rather unnatural on the lattice, 
have been chosen to simplify the proof and could be omitted at the expense of 
using the  stationary phase method to evaluate the integral over the manifold
$\nu(q)=\omega$ in (\ref{eq-assumptionbis}), instead of using (\ref{eq-f(x,t)_as_Fourier_transfo}).
In contrast, the first condition $d \geq 2$ is crucial:
in dimension $d=1$, (\ref{eq-integrability_assumption}) is 
not fulfilled.  To see this, let us assume that (ii-iii) hold and that 
the baths are initially at thermal equilibrium, i.e., 
$\zeta (q) = ( \e^{\beta |q|}-1)^{-1}$. For $d=1$,  (\ref{eq-assumptionbis}) gives
\be
f(x,t) 
 = \int_{-r}^r \frac{\D q}{2\pi} \, 
  |g_0(q)|^2 \frac{\e^{i q (t+x)} + \e^{i q (t-x)}}{|\e^{\beta q} - 1 |}\,.
\ee
It is clear that $\sup_{x\in \integer} | f(x,t)|$ does not decay to zero at large times $t$. As a result, 
the integral in  (\ref{eq-integrability_assumption}) diverges.

\vspace{2mm}

%%%%%%%%%%%%%%%%%%%%%%%%%%%%%%%%%%%%%%%%%%%

\end{document}